\documentclass[twocolumn]{aastex6}
\usepackage{graphicx}
\usepackage{amssymb}
\usepackage{amsmath}
\usepackage{hyperref}

\def\gtrsim{\mathrel{\hbox{\rlap{\hbox{\lower4pt\hbox{$\sim$}}}\hbox{$>$}}}}
\def\lesssim{\mathrel{\hbox{\rlap{\hbox{\lower4pt\hbox{$\sim$}}}\hbox{$<$}}}}
\def\gtrsim{\mathrel{\hbox{\rlap{\hbox{\lower4pt\hbox{$\sim$}}}\hbox{$>$}}}}
\def\farcs{\hbox{$.\!\!^{\prime\prime}$}}
\def\farcm{\hbox{$.\!\!^{\prime}$}}

\begin{document}
\title{Chandra Observations of the field containing HESS J1616--508  }
\author{Jeremy Hare\altaffilmark{1}, Oleg Kargaltsev\altaffilmark{1}, George G. Pavlov\altaffilmark{2}, Blagoy Rangelov\altaffilmark{1,3}, Igor Volkov\altaffilmark{1,4} }
\altaffiltext{1}{Department of Physics, The George Washington University, 725 21st St. NW, Washington, DC 20052}
\altaffiltext{2}{Department of Astronomy $\&$ Astrophysics, Pennsylvania State University, 525 Davey Lab, University Park, PA 16802, USA}
\altaffiltext{3}{Department of Physics, Texas State University, 749 N. Comanche St., San Marcos, TX 78666}
\altaffiltext{4}{University of Maryland, College Park, MD 20742, USA}
\email{jeh86@gwu.edu}

\begin{abstract}
We report the results of three {\sl Chandra} observations covering most of the extent of the TeV $\gamma$-ray source HESS J1616--508 and a search for a lower energy counterpart to this source. We detect 56 X-ray sources, of which 37 have counterparts at lower frequencies, including a young massive star cluster, but none of them appears to be a particularly promising counterpart to the TeV source. The brightest X-ray source, CXOU J161423.4--505738 with a flux $F_{\rm 0.5-7kev}\approx5\times10^{-13}$ erg cm$^{-2}$ s$^{-1}$, has a hard spectrum that is well fit by a power-law model with a photon index $\Gamma=0.2\pm0.3$ and is a likely intermediate polar CV candidate. No counterparts of this source were detected at other wavelengths. CVs are not known to produce extended TeV emission, and the source is also largely offset $(19')$ from HESS J1616--508, making them unlikely to be associated. We have also set an upper limit on the X-ray flux of PSR J1614--5048 in the 0.5-8 keV band ($F_{\rm 0.5-8 keV}<5\times10^{-15}$ erg cm$^{-2}$ s$^{-1}$ at a 90\% confidence level). This makes PSR J1614--5048 one of the least X-ray efficient pulsars known, with an X-ray efficiency $\eta_{\rm 0.5-8keV}=L_{\rm 0.5-8kev}/\dot{E} < 2\times10^{-5}$. We find no evidence supporting the association between the pulsar and the TeV source. We rule out a number of X-ray sources as possible counterparts to the TeV emission and do not find a plausible counterpart among the other sources. Lastly, we discuss the possible relation of PSR J1617--5055 to HESS J1616--508 in light of the new observations.
\end{abstract}

\section{Introduction}

The High Energy Stereoscopic System (H.E.S.S.) is an array of atmospheric Cherenkov telescopes sensitive to photons above 100 GeV, which has discovered $>$ 75 TeV sources in the Galactic plane \citep{2005Sci...307.1938A}\footnote{See \url{https://www.mpi-hd.mpg.de/hfm/HESS/pages/home/som/2016/01/}.}. Among these sources there are a few that have not been associated with any known objects (\citeauthor{2008A&A...477..353A} \citeyear{2008A&A...477..353A}; \citeauthor{2013arXiv1305.2552K} \citeyear{2013arXiv1305.2552K}). The most straightforward way to understand what type of source is producing the high energy emission is to use observations at longer wavelengths. However, for some of the TeV sources, attempts to find their X-ray counterparts  with {\sl Chandra} or {\sl XMM-Newton} either failed or produced an ambiguous result, causing these sources to be dubbed ``{\sl dark accelerators}'' \citep{2005Sci...307.1938A}. Observations of these fields with {\sl Fermi-LAT} have revealed possible GeV counterparts for many TeV sources (\citeauthor{2012ApJS..199...31N} \citeyear{2012ApJS..199...31N}; \citeauthor{2013ApJ...773...77A} \citeyear{2013ApJ...773...77A}; \citeauthor{2015ApJS..218...23A} \citeyear{2015ApJS..218...23A}), but the angular resolution of the LAT is often not sufficient to establish a firm association. 

Spatially extended TeV sources are often associated with energetic pulsars located nearby (10$'$$-$15$'$ from the center of the TeV emission; see \citeauthor{2012ASPC..466..167K} \citeyear{2012ASPC..466..167K} and \citeauthor{2013arXiv1305.2552K} \citeyear{2013arXiv1305.2552K} for reviews).  Pulsars eject relativistic particles into their surroundings and can fill large bubbles (plerions) with these particles during their lifetime. These particles can produce TeV photons by inverse Compton (IC) upscattering of background photons. The large observed offsets between pulsar position and TeV source could, in some cases, be explained by the {\sl crushed plerion scenario}, which assumes that the asymmetric  reverse shock from the host supernova remnant (SNR) has displaced the plerion from the pulsar \citep{2001ApJ...563..806B}. The offset extended TeV source would then represent a {\sl relic plerion} filled with aged electrons. These same pulsars can also power compact X-ray pulsar wind nebulae (PWNe), whose synchrotron emission is produced by ``younger,'' more energetic electrons. In some cases, such nebulae are detected before the pulsar is discovered (see Table 1 in \citeauthor{2013arXiv1305.2552K} \citeyear{2013arXiv1305.2552K}) revealing the nature of the TeV source.

HESS J1616--508 (hereafter HESS J1616) is one of the brightest unidentified extended ($\sim16'$ rms size) TeV sources \citep{2006ApJ...636..777A}. The source has a 1$-$10 TeV flux $F_{\rm 1-10 TeV}\approx1.7\times10^{-11}$ erg s$^{-1}$ cm$^{-2}$, and its spectrum can be fit by a power-law model with $\Gamma=2.34\pm0.06$. HESS J1616's center is located about $10'$ east of the young ($\tau\equiv P/2\dot{P}=8.13$ kyr) and energetic ($\dot{E}=1.6\times10^{37}$ erg s$^{-1}$) pulsar, PSR J1617--5055 (PSR J1617 hereafter; \citeauthor{1998ApJ...503L.161K} \citeyear{1998ApJ...503L.161K}). 
PSR J1617 has a period $P=69$ ms, magnetic field $B=3.1\times10^{12}$ G, and is estimated to be at a distance of 6.5 kpc. This pulsar is energetic enough to power the TeV source, however, no other links between the PSR J1617 and HESS J1616 have been found (\citeauthor{2009ApJ...690..891K} \citeyear{2009ApJ...690..891K}; K+09 hereafter).

The field of HESS J1616 has been observed in X-rays multiple times. \cite{2007PASJ...59S.199M} observed the center of the HESS J1616 field with {\sl Suzaku} XIS but found no counterpart to a limiting flux of 3$\times10^{-13}$ erg s$^{-1}$ cm$^{-2}$ in the 2$-$10 keV energy range. The field was also observed with {\sl INTEGRAL}/IBIS/ISGRI and {\sl Swift} XRT. The data were analyzed by \cite{2007MNRAS.380..926L}; however, due to the lack of any other obvious X-ray counterpart, they concluded that the most likely counterpart to the TeV source was PSR J1617. 

K+09 analyzed a dedicated 60 ks {\sl Chandra X-ray Observatory} ({\sl CXO})  observation of PSR J1617, as well as the available archival data (including {\sl XMM-Newton} and {\sl Chandra} observations of nearby SNRs). K+09 found that the under-luminous X-ray PWN ($L_{\rm 0.5-8 keV,PWN}\sim 3.5\times10^{33}$ erg s$^{-1} \sim0.2L_{\rm 0.5-8 keV, PSR} \sim 2\times10^{-4}\dot{E}$, at $d = 6.5$ kpc) does not extend in the direction toward the center of the TeV source (unlike, e.g., the X-ray PWN of PSR B1823--13 and HESS J1825--137; \citeauthor{2008ApJ...675..683P} \citeyear{2008ApJ...675..683P}). K+09 suggested that even if PSR J1617 is responsible for a fraction of the TeV emission, it could be that HESS J1616 is a double or multiple source, which might include an unknown SNR or PWN. K+09 also reported an X-ray source, dubbed ``Source X'', with a flux $F_{\rm 0.5-8 keV}\sim2\times10^{-13}$ erg cm$^{-2}$ s$^{-1}$ in the $0.5-8$ keV band located close to the center of the TeV source. However, this source was imaged 16$'$ off-axis (in an archival {\sl Chandra} observation), making it impossible to distinguish between a single extended source or several point sources.

There is also an extended ($\sigma\sim19'$) source reported at the exact same coordinates in both the {\sl Fermi-LAT} 3FGL and 2FHL source catalogs (\citeauthor{2015ApJS..218...23A} \citeyear{2015ApJS..218...23A}; \citeauthor{2016ApJS..222....5A} \citeyear{2016ApJS..222....5A}), named 3FGL J1616.2--5054e and 2FHL J1616.2--5054e, respectively. This source is spatially coincident with the TeV source (i.e., the center of this {\sl Fermi-LAT} source is only $\sim 36''$ from the center of HESS J1616). The LAT source has fluxes $F=(1.56\pm0.08)\times10^{-10}$ erg cm$^{-2}$ s$^{-1}$ in the 1$-$100 GeV band, $F=(7.3\pm1.7)\times10^{-11}$ erg cm$^{-2}$ s$^{-1}$ in the 50 GeV$-$2 TeV band and spectral indices $\Gamma=2.14\pm0.03$, $\Gamma=1.74\pm0.18$ in the 3FGL and 2FHL catalogs, respectively. The spectrum of this source shows no sign of curvature, which is typically seen in the {\sl Fermi-LAT} spectra of pulsars\footnote{The source is also listed as being variable in the 3FGL catalog, but this is solely due to a processing issue that occurs with extended sources described in Section 3.6 of \cite{2015ApJS..218...23A}.} \citep{2013ApJS..208...17A}. 

Another potential counterpart of HESS J1616 is PSR J1614--5048 (PSR J1614 hereafter; \citeauthor{1992MNRAS.255..401J} \citeyear{1992MNRAS.255..401J}), located $22'$ west of the TeV source center. PSR J1614 is a young pulsar, based on its spin-down age $\tau=7.4$ kyr. However, in some cases, this age may be a poor indicator of the true age. The pulsar is estimated to be at $d\sim7$ kpc determined from its dispersion measure DM$=586$ pc cm$^{-3}$ \citep{1993ApJ...411..674T}. The pulsar has a period $P=232$ ms, a rather low spin-down power for a young pulsar $\dot{E}=1.6\times10^{36}$ erg s$^{-1}$,  and a high surface dipole magnetic field $B=1.1\times10^{13}$ G \citep{2000MNRAS.317..843W}. This pulsar is also interesting because it has experienced two of the largest glitches ever observed and is also one of the noisiest pulsars (\citeauthor{2000MNRAS.317..843W} \citeyear{2000MNRAS.317..843W}; \citeauthor{2013MNRAS.429..688Y} \citeyear{2013MNRAS.429..688Y}).

Here we report the results of our analysis of three new {\sl Chandra} observations, two of them cover the center of HESS J1616 (ObsIDs 16956 and 17676) and another one (ObsID 16955) covers the field of PSR J1614 (Section \ref{data}). We also search for multi-wavelength (MW) counterparts to the detected X-ray sources (Section \ref{counter}),  use the MW data to classify the X-ray sources detected in these observations, and discuss whether any of these sources could be a counterpart to HESS J1616 (Section \ref{result}). We find that PSR J1614 is not detected in X-rays (Section \ref{J1614}) and summarize our findings in Section \ref{conc}.

\begin{figure*}
\begin{center}
\includegraphics[scale=0.873,trim=0 5 0 0]{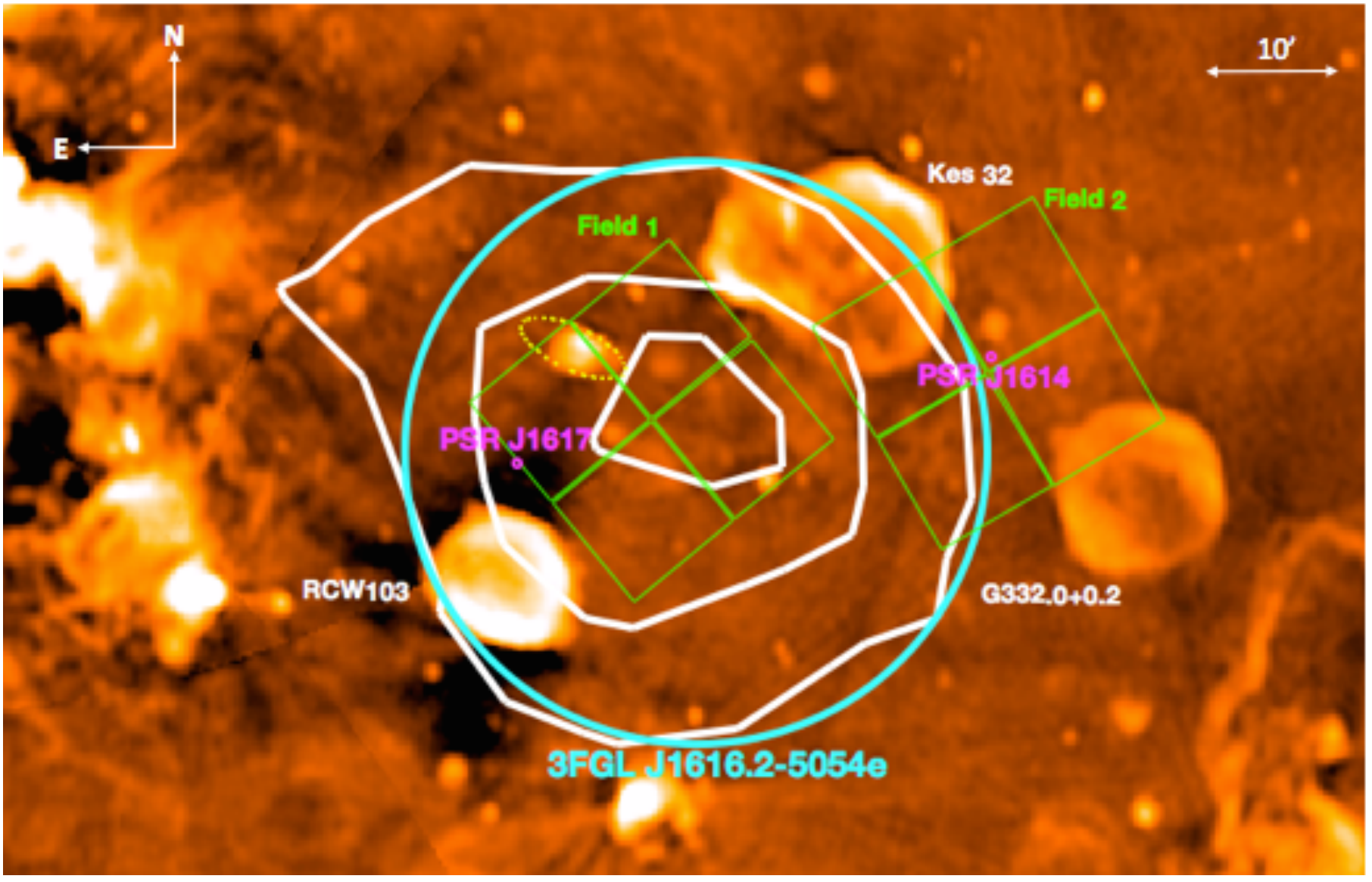}
\includegraphics[scale=0.863, trim=0 0 0 0]{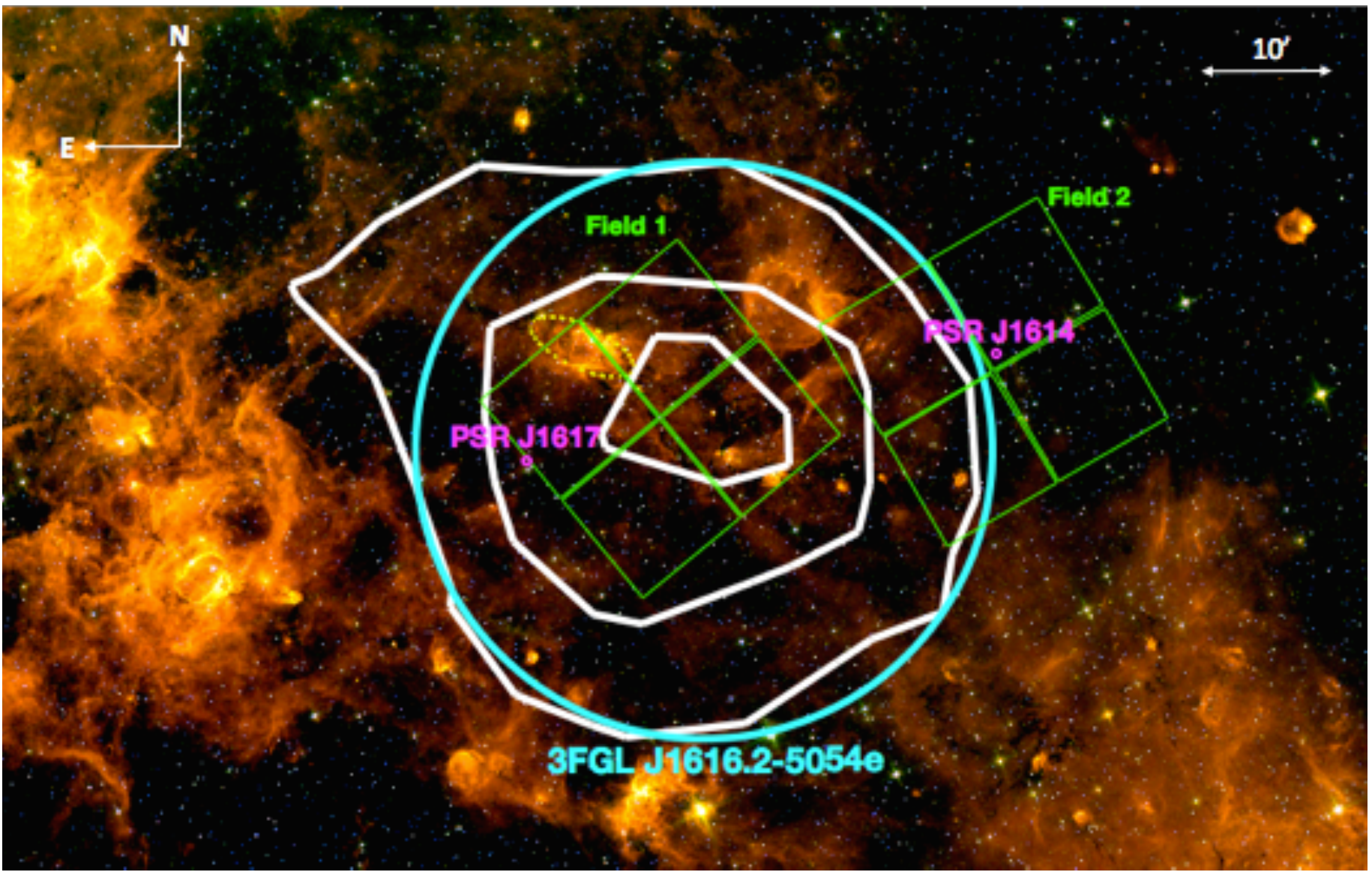}
\caption{{\sl Top}:  The 843 MHz radio image of the field of HESS J1616 from the Sydney University Molonglo Sky Survey (SUMSS; \citeauthor{2003MNRAS.342.1117M} \citeyear{2003MNRAS.342.1117M}). The ACIS-I fields of view from the {\sl CXO} observations are shown in green (two of the observations land on top of each other comprising Field 1). The HESS J1616 contours are shown in white, the extent of the {\sl Fermi-LAT} source 3FGL 1616.2$-$5054e is shown in cyan, and the two known pulsars in this field are shown in magenta. The young star cluster discussed in Section \ref{counter} is shown as a dashed yellow ellipse. Three supernova remnants (RCW103, Kes 32, and G332.0$+$0.2) can be seen in the radio image. {\sl Bottom}: Spitzer IRAC false color image (red--8.0 $\mu$m; green--5.8 $\mu$m; blue--3.6 $\mu$m) of the same part of the sky.}
\label{MWfig}
\end{center}
\end{figure*}

\section{{\sl Chandra} Observations and Data Reduction}
\label{data}

We have analyzed three {\sl Chandra} observations. The first two, ObsIDs 16956 and 17676 observed on 2015 June 18 and 19, covered the central region of HESS J1616 (Field 1 hereafter; 24.7 ks and  33.6 ks exposures, respectively), while the third observation, ObsID 16955 observed on 2015 June 15, covered the field containing PSR J1614 (Field 2, hereafter; 59.3 ks exposure). Fields 1 and 2 are shown in Figure \ref{MWfig}. All three observations were obtained with the ACIS-I instrument operated in timed exposure mode using the Very Faint telemetry format. The data were processed with the {\sl Chandra} Interactive Analysis of Observations (CIAO\footnote{\url{http://cxc.harvard.edu/ciao/}}) software version 4.6 and the {\sl Chandra} Calibration Database (CALDB) version 4.6.3. We restricted our analysis  to the 0.5--7 keV range, unless otherwise specified. The spectra were fit using XSPEC version 12.8.2.

\begin{figure*}
\begin{center}
\includegraphics[scale=0.85,trim=0 0 0 0]{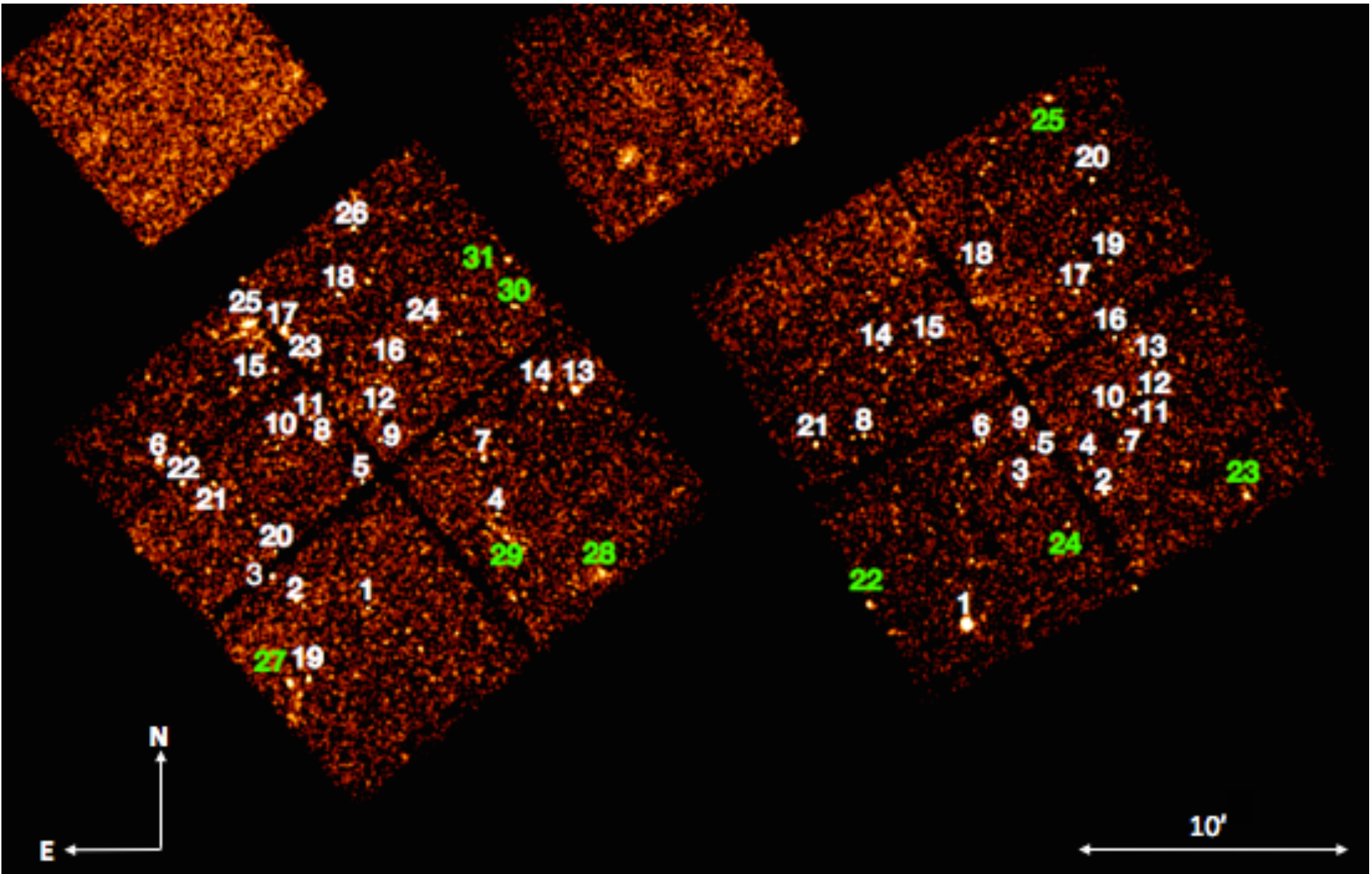}
\caption{ Field 1 (ObsIDs 16956 and 17676{\bf ; Left}) and Field 2 (ObsID 16955{\bf; Right}) {\sl CXO} images. The sources labelled in white are those found by {\tt wavdetect} with a source significance greater than 6. The sources labelled in green are those that were manually added (see Section \ref{data}). The images are binned by a factor of 4 (pixel size $2''$) and smoothed by a Gaussian kernel with a width of 3$''$.  }
\label{cxofig}
\end{center}
\end{figure*}

CIAO's source detection routine {\tt wavdetect} \citep{2002ApJS..138..185F} was used to detect X-ray point sources and determine their coordinates in the {\sl Chandra} images (see Table \ref{tab1} and Figure \ref{cxofig}). In order to detect fainter point sources, the data for ObsIDs 17676 and 16956 (Field 1) were merged\footnote{Standard CIAO procedures found at \url{http://cxc.harvard.edu/ciao/threads/wavdetect_merged/} were followed to merge the data. We used an exposure time weighted average PSF map in the calculation of the merged PSF.}  prior to running {\tt wavdetect}. The {\tt srcflux} CIAO tool was then run individually on each observation (using the coordinates found by running {\tt wavdetect} on Field 1) and the fluxes were averaged by weighting the flux in each observation with its corresponding exposure time. There were 26 and 21 sources detected with a {\tt wavdetect} reported  significance $>$ 6 for Field 1 and Field 2, respectively (see Figure \ref{cxofig}). A cut on significance is needed to ensure that meaningful hardness ratio values can be obtained. Several off-axis sources (5 in Field 1 and 4 in Field 2) had {\tt wavdetect} significance values $<6$ due to a higher background in their immediate vicinity caused by nearby very faint sources or hints of extended emission. We have added these sources to our source list by hand and marked them with green labels in Figure \ref{cxofig}.

The X-ray spectra for the five brightest sources ($\geq$ 100 counts) have been created by the {\tt srcflux} CIAO tool. For sources with $>$ 200 counts, we have binned the X-ray spectra to have a minimum of 10 counts per bin. Otherwise, the spectra were not binned and were fit using C-statistics \citep{1979ApJ...228..939C}. For Field 1, the spectra were extracted separately for each observation and combined using the {\sl combine\_spectra} CIAO tool\footnote{We have verified that simultaneous fitting produces the same results (within errors) in this particular case.}. All fits used the XSPEC {\tt tbabs} model \citep{2000ApJ...542..914W} for interstellar absorption.

\begin{deluxetable*}{ccccccccc}
\tablecaption{{\sl CXO} sources detected in both Field 1 and Field 2.}
\tablewidth{0pt}
\tablehead{
\colhead{Field} & \colhead{Source \#} & \colhead{CXOU\tablenotemark{a}} & \colhead{R.A.} & \colhead{Decl.} & \colhead{Radius\tablenotemark{b}} & \colhead{Counts\tablenotemark{c}} & \colhead{$F$\tablenotemark{d}} & \colhead{Class\tablenotemark{e} (\%)}}
\startdata
1 & 1 & J161637.3--505705 & 244.15530 & $-$50.95141 & 1.4 & $25\pm7$ & 1.0(4) &   ? \\
1 & 2 & J161653.3--505647 & 244.22213 & $-$50.94635 & 1.2 & $41\pm10$ & 0.9(4) &  YSO (72) \\
1 & 3 & J161658.6--505558 & 244.24418 & $-$50.93278 & 0.9 & $67\pm12$ & 4(1) & ? \\
1 & 4 & J161608.3--505348 & 244.03436 & $-$50.89669 & 1.3 & $22\pm6$ & 0.8(4) &  AGN (100) \\ 
1 & 5 & J161638.4--505240 & 244.15991 & $-$50.87771 & 0.6 & $29\pm7$ &  1.6(8)  & AGN (100)\\
1 & 6 & J161723.7--505152 & 244.34871 & $-$50.86454 & 0.8 & $184\pm20$ & 5.3(9) & STAR (99)\tablenotemark{f} \\
1 & 7 & J161611.5--505152 & 244.04781 & $-$50.86456  & 0.8 & $44\pm10$ & 1.5(6) &  ?\\
1 & 8 & J161647.1--505119 & 244.19630 & $-$50.85513 & 0.8 & $14\pm5$ & 0.3(2) &  ?\\
1 & 9 & J161634.3--505110 &  244.14304 & $-$50.85288 & 0.5 & $25\pm7$ & 1.6(8) &  ?\\ 
1 &10 & J161656.1--505107 & 244.23365 & $-$50.85182 & 1.0 & $18\pm6$ & 0.4(2) &  ?\\
1 &11 & J161650.0--505025 & 244.20819 & $-$50.84026 & 0.7 & $36\pm8$ & 2(1) &  AGN (100)\\
1 &12 & J161634.4--505015 & 244.14317 & $-$50.83755 & 0.5 & $60\pm11$ &  1.6(5) & STAR (99) \\
1 &13 & J161551.0--504925 & 243.96233  & $-$50.82370 & 0.7 & $388\pm28$ &  12(1) & STAR (72)\\
1 &14 & J161557.8--504921 & 243.99093 & $-$50.82259 & 1.1 & $72\pm12$ & 2.5(7) &  STAR (81) \\
1 &15 & J161657.6--504842 & 244.24010 & $-$50.81172 & 0.7 & $82\pm11$ & 1.9(5) &  ?\\
1 &16 & J161632.2--504837 & 244.13400 & $-$50.81039 & 1.0 & $17\pm6$ & 0.3(2) &  STAR (99)\\
1 &17 & J161655.6--504713 & 244.23184 & $-$50.78698 & 1.1 & $41\pm9$ & 1.5(6) &  AGN (100)\\
1 &18 & J161643.5--504604 & 244.18131 & $-$50.76780 & 0.8 & $116\pm16$ &  5(1) & ?\\
1 &19 & J161650.5--505935 & 244.21025 & $-$50.99296 & 2.0 & $51\pm12$ & 0.9(4) &  STAR (78) \\
1 &20 & J161657.8--505505 & 244.24097 & $-$50.91803 & 1.3 & $21\pm7$ & 0.5(3) &  ? \\ 
1 &21 & J161711.6--505242 & 244.29837 & $-$50.87837 & 1.4 & $28\pm8$ & 0.6(3) &  ?\\ 
1 &22 & J161718.5--505241 & 244.32714 & $-$50.87801 & 1.4 & $50\pm11$ & 2.2(8) &  ?\\
1 &23 & J161655.8--504722 & 244.23247 & $-$50.78940 & 0.8 & $82\pm13$ & 3.1(9) &  ? \\
1 &24 & J161624.4--504714 & 244.10171  & $-$50.78713 & 1.4 & $19\pm5$ & 0.7(3) &  AGN (100)\\
1 &25 & J161703.6--504705 & 244.26489 & $-$50.78476 & 0.9 & $100\pm15$ & 3.8(9) &  YSO (80) \\
1 &26 & J161640.4--504344 & 244.16843 & $-$50.72899 & 1.8 & $57\pm12$ & 1.1(4) &  STAR (75) \\
2 & 1 & J161423.4--505738 & 243.59749 & $-$50.96054 & 0.9 & $663\pm26$ &  47(3) &  ? \\
2 & 2 & J161353.0--505259 & 243.47086 & $-$50.88308 & 1.1 & $50\pm7$ & 1.5(4) &   STAR (73) \\
2 & 3 & J161411.5--505245 & 243.54776 & $-$50.87906 & 0.8 & $59\pm8$ & 1.0(2) &  YSO (70)\\
2 & 4 & J161356.1--505156 & 243.48378 & $-$50.86546 & 1.2 & $24\pm5$ & 1.9(6) &  AGN (100) \\ 
2 & 5 & J161409.0--505124 & 243.53754 & $-$50.85665 & 0.7 & $38\pm6$ & 0.8(2) &  STAR (75)\\
2 & 6 & J161420.2--505110 & 243.58398 & $-$50.85288 & 0.8 & $24\pm5$ & 0.5(1) &   YSO (76)\\
2 & 7 & J161349.3--505110 & 243.45531 & $-$50.85279  & 1.2 & $30\pm6$ &  0.8(2) &  STAR (81)\\
2 & 8 & J161446.5--505102 & 243.69362 & $-$50.85054 & 1.3 & $27\pm5$ & 1.6(5) &  ?\\
2 & 9 &  J161411.1--505058 & 243.54634 & $-$50.84936 & 0.9 & $16\pm4$ & 1.2(4) &  ?\\ 
2 &10 & J161350.7--505014 & 243.46142 & $-$50.83718 & 1.0 & $27\pm5$ & 0.5(2) & STAR (88)\\
2 &11 & J161346.3--505008 & 243.44293 & $-$50.83566 & 0.8 & $87\pm9$ & 2.0(4) &  STAR (95)\\
2 &12 & J161345.4--504928 & 243.43905 & $-$50.82436 & 1.2 & $25\pm5$ & 0.18(6) &  STAR (89)\\
2 &13 & J161342.2--504824 & 243.42573 & $-$50.80661 & 0.8 & $84\pm9$ & 3.3(6) &  YSO (90)\\
2 &14 & J161443.0--504759 & 243.67929 & $-$50.79973 & 0.7 & $79\pm9$ & 1.9(4) &  ?\\
2 &15 & J161432.4--504744 & 243.63479 & $-$50.79554 & 0.9 & $17\pm4$ &0.9(3)  & AGN (100)\\
2 &16 & J161351.1--504732 & 243.46269 & $-$50.79212 & 1.0 & $27\pm5$ & 1.0(3) &  AGN (100)\\
2 &17 & J161359.5--504555 & 243.49771 & $-$50.76525 & 0.8 & $33\pm6$ & 1.8(5) &   AGN (100)\\
2 &18 & J161421.2--504511 & 243.58837 & $-$50.75317 & 0.7 & $54\pm7$ & 3.2(7) &  AGN (100)\\
2 &19 & J161352.1--504453 & 243.46697 & $-$50.74800 & 1.3 & $25\pm5$ & 0.9(3) &  STAR (78) \\
2 &20 & J161356.2--504158 & 243.48395 & $-$50.69956 & 1.4 & $61\pm8$ & 4(1) &   STAR (86)\\ 
2 &21 & J161457.3--505119 & 243.73884 & $-$50.85535 & 1.7 & $40\pm7$ & 2.5(7) &  ?\\ 
\hline
1 &27 & J161654.6--505943 & 244.22763 & $-$50.99525 & 1.8 & $75\pm14$ & 3(1) &  STAR (71) \\
1 &28 & J161544.9--505551 & 243.93705 & $-$50.93097 & 2.4 & $55\pm12$ & 0.8(3) &  AGN (100) \\
1 &29 & J161606.2--505437 & 244.02600 & $-$50.91042 & 1.5 & $22\pm7$ & 0.8(5) &  AGN (100) \\
1 &30 & J161604.2--504630 & 244.01771 & $-$50.77508 & 2.0 & $28\pm9$ & 2(1) &   STAR (85) \\
1 &31 & J161605.9--504450 & 244.02439 & $-$50.74739 & 2.1 & $46\pm10$ &  46(14) & ? \\
2 &22 & J161445.3--505657 & 243.68883 & $-$50.94925 & 2.7 & $58\pm9$ & 6(1) &  STAR (79)\\ 
2 &23 & J161321.3--505303 & 243.33866 & $-$50.88428 & 2.6 & $66\pm9$ & 7(2) &  ?\\ 
2 &24 & J161401.0--505406 & 243.50411 & $-$50.90169 & 1.6 & $27\pm6$ & 2.3(7) &  ?\\ 
2 &25 & J161405.9--503908 & 243.52452 & $-$50.65225 & 2.5  & $59\pm9$ & 8(2) &  ?\\ 
\label{tab1}
\enddata
\tablecomments{Sources at the bottom of the table, below the horizontal line, have {\tt wavdetect} significances $<6$ and were added in by hand (see Section \ref{data}).}
\tablenotetext{\rm a}{Source name following the standard {\sl Chandra} naming convention. }
\tablenotetext{\rm b}{The 2$\sigma$ positional uncertainty radius used for the MW counterpart search for each X-ray source (see Section \ref{counter}).}
\tablenotetext{\rm c}{Number of background-subtracted counts in the $0.5-7$~keV range in the regions used by {\tt srcflux} (see Section \ref{result}).}
\tablenotetext{\rm d}{Observed fluxes in the $0.5-7$~keV range in units of $10^{-14}$ erg~s$^{-1}$~cm$^{-2}$. The fluxes for the sources in Field 1 were extracted separately from each observation and combined using an exposure weighted average. The error shown in parentheses is the 90\% error on the final digit.}
\tablenotetext{\rm e}{Classification and its confidence according to the automated classification pipeline (see Section \ref{counter}). ``?" denotes cases with less confident ($<$ 70\% confidence) classifications.}
\tablenotetext{\rm f}{This source was classified using MW information from a source just outside of the 2$\sigma$ error circle. See Field 1 Source 6 in Section \ref{result} for more details.}
\end{deluxetable*}

\section{MW Counterparts to X-ray Sources}
\label{counter}

Along with calculating X-ray properties for each source (i.e., the 0.5$-$2 keV flux, 2$-$7 keV flux, and two hardness ratios), we have also searched MW catalogs for potential counterparts\footnote{ We attempted to correct the absolute astrometry (following the procedure described in \url{http://cxc.harvard.edu/ciao/threads/reproject_aspect/}) of the ACIS images by cross-matching the detected X-ray sources with the 2MASS catalog sources \citep{2006AJ....131.1163S} in each field. However, the resulting corrections were insignificant. They did not impact the choice of MW counterparts for any of the X-ray sources.} to these sources and compiled a set of MW parameters for each of the sources with counterparts. The MW photometry was taken in optical from the USNO-B \citep{2003AJ....125..984M} catalog, the near-infrared (NIR) from the Two Micron All Sky Survey (2MASS; \citeauthor{2006AJ....131.1163S} \citeyear{2006AJ....131.1163S}), and in the IR from the {\sl Wide-field Infrared Survey Explorer} ({\sl WISE}; \citeauthor{2012wise.rept....1C} \citeyear{2012wise.rept....1C}). We found that 16 out of 31 and 14 out of 25 of the X-ray sources had MW counterparts in Field 1 and Field 2, respectively. MW sources were considered potential counterparts if they were within the error radius of each X-ray source. This error radius was calculated using the empirical relations provided by the {\sl Chandra} Multi-wavelength Project (see Equation 12 in \citeauthor{2007ApJS..169..401K} \citeyear{2007ApJS..169..401K}). 
 Then, we followed the {\sl Chandra} Source Catalog \citep{2010ApJS..189...37E}, and added an additional 2$\sigma$ astrometric error of $0\farcs39$ in quadrature (see Table \ref{tab1}). The average 2$\sigma$ error circle for all of the sources in both fields is 1$\farcs2$. The chance coincidence probabilities for each catalog were calculated by finding the average source density in these fields, $\rho$, and calculating the probability of having one or more sources within a randomly placed circle, $P=1-\exp{(-\rho\pi r^2)}$. The source densities for each of the catalogs in these fields are $\rho_{\rm USNOB}=0.00342$ arcsec$^{-2}$, $\rho_{\rm 2MASS}=0.00895$ arcsec$^{-2}$, and $\rho_{\rm WISE}=0.00168$ arcsec$^{-2}$, leading to chance coincidence probabilities of $1.5\%$, $4.0\%$, and $0.8\%$, respectively, for $r=1\farcs2$. The 2MASS catalog has the largest chance coincidence probability (of the cross-matched catalogs), which implies that up to $\sim$2 2MASS sources may be spurious cross-matches with an X-ray source for both fields combined. None of the X-ray sources had more than one {\sl WISE} or 2MASS counterpart; however, two X-ray sources (19 and 26 in Field 2) had two USNO-B counterparts. For these cases, the closest optical source to the X-ray source position was taken to be the optical counterpart. The MW magnitudes for each counterpart are given in Table \ref{17MW}.

The magnitudes\footnote{All magnitudes for the USNO-B, 2MASS, and {\sl WISE} catalogs are in the Vega-magnitude system.} for the MW counterparts together with the X-ray properties (fluxes in two bands and two hardness-ratios), add up to 19 features (or parameters), which are then used to classify these sources with our machine learning pipeline (see the Appendix of \citeauthor{2016ApJ...816...52H} \citeyear{2016ApJ...816...52H}). One minor change to the pipeline used in \cite{2016ApJ...816...52H}  is that we no longer use the See5 implementation of the C5 decision tree algorithm \citep{1993cpml.book.....Q} and solely rely on a Random Forest classifier \citep{breiman2001} implemented in python\footnote{\url{http://scikit-learn.org/stable/modules/generated/sklearn.ensemble.RandomForestClassifier.html}} \citep{2012arXiv1201.0490P}. We have also reddened the AGNs in the training dataset using the total Galactic absorption column in the direction of our observations ($N_{H}=2.4\times10^{22}$ cm$^{-2}$; \citeauthor{1990ARA&A..28..215D} \citeyear{1990ARA&A..28..215D}) to make them similar to the background AGNs in this field.

To search for possible radio counterparts we used radio data from The Sydney University Molonglo Sky Survey (SUMMS; \citeauthor{2003MNRAS.342.1117M} \citeyear{2003MNRAS.342.1117M}) taken at 843 MHz, which is shown in the top panel of Figure \ref{MWfig}. The bottom panel of Figure \ref{MWfig} shows a false color (red--8.0  $\mu$m green--5.8 $\mu$m blue--3.6 $\mu$m) {\sl Spitzer} IRAC image. There is a bright radio source seen towards the top of CXO Field 1, which can also be seen in the {\sl Spitzer} image. This is a known young massive stellar \citep{2004AJ....127.2817R} with a lower limit on the bolometric luminosity of $8.7\times10^{4}$ $L_{\odot}$, it contains a number of O and B type stars (35 cluster member candidates detected down to the completeness limit). Field 2 partially covers the SNRs Kes 32 and G332.0+0.2, with Kes 32 partially seen in the {\sl CXO} image. G332.0+0.2 is not detected in the {\sl Chandra} observation. 

Lastly, we have also searched the {\sl INTEGRAL/IBIS} 7-year source catalog \citep{2010A&A...523A..61K}, which has been observing the field of HESS J1616 regularly since 2013 February. We found only one source (associated with the nearby pulsar PSR J1617) within $30'$ of the center of each field. Therefore, we conclude that there are no {\sl INTEGRAL} counterparts to any other sources in these fields down to a limiting flux of 3.7$\times10^{-12}$ erg s$^{-1}$ cm$^{-2}$ in the 17-60 keV energy band \citep{2010A&A...523A..61K}.

\begin{deluxetable*}{ccccccccccc}
\tablecaption{ Magnitudes of MW counterparts to {\sl Chandra} sources in Field 1 and Field 2 (see Figure \ref{cxofig} and Section \ref{counter}).}
\tablewidth{0pt}
\tablehead{
\colhead{Field} & \colhead{Source \#} & \colhead{$B$\tablenotemark{a}} & \colhead{$R$\tablenotemark{a}} & \colhead{$I$\tablenotemark{a}} & \colhead{$J$\tablenotemark{b}} & \colhead{$H$\tablenotemark{b}} & \colhead{$K$\tablenotemark{b}} & \colhead{$W1$\tablenotemark{c}} &\colhead{$W2$\tablenotemark{c}} & \colhead{$W3$\tablenotemark{c}}}
\startdata
 1 & 1 & ... & ... & ... & 15.62 & 14.41 & 13.79 & ... & ... & ...   \\
 1 & 2 & ... & ... & ...  & 13.21 & 12.68 & 12.49 & ... & ... & ...   \\
 1 & 6\tablenotemark{d} & 10.38 & 9.77 & 9.56 & 8.92 & 8.713 & 8.66 & 8.51 & 8.56 & 8.32 \\
 1 & 7  & ... & ...  & ... & 14.19 & 11.83 & 10.06 & 7.91 & 6.94 & 3.26  \\
 1 & 10  & 19.97 & 17.68  & 18.02 & ... & ... & ... & ... & ... & ...  \\
 1 & 12  & ... & ...  & ... & 10.59 & 9.88 & 9.39 & 8.36 & 7.92 & 6.54  \\
 1 & 13  & 14.42 & 11.81  & 11.01 & 9.88 & 9.21 & 8.99 & 8.81 & 8.84 & 7.57  \\
 1 & 14  & 15.40 & 13.67  & 12.86 & ... & ... & ... & ... & ... & ...  \\
 1 &  16  & 17.59 & 15.59 & 14.70 & 14.22 & 13.60 & 13.42 & ... & ... & ...  \\
 1 &  19\tablenotemark{e} & 14.80 & 14.18 & 13.37 & 12.15 & 11.74 & 11.62 & 11.44 & 11.39 & 7.02  \\
 1 &   20  & ... & ... & ... & 13.63 & 13.07 & 12.95 & ... & ... & ...  \\
 1 &   21  & ... & ... & .... & 13.58 & 12.12 & 11.49 & ... & ... & ...  \\
 1 &   22  & ... & ... & ... & 12.86 & 10.99 & 10.18 & 9.72 & 9.66 & 9.23  \\
 1 &   25 & ... & ... & ... & 14.17 & 12.59 & 11.59 & ... & ... & ...  \\
 1 &   26\tablenotemark{e} & 14.82 & 13.73 & 13.09 & 12.80 & 12.35 & 12.16 & 12.44 & 12.92 & 7.78 \\
 2 & 2 & 18.63 & 16.25 & 15.84 & ... & ... & ... & ... & ... & ...   \\
 2 &  3 & ... & ... & ...  & 11.84 & 11.14 & 10.96 & ... & ... & ...   \\
 2 &  5 & 17.49 & 14.82 & ...  & ... & ... & ... & ... & ... & ...   \\
 2 &  6  & ... & ...  & ... & 13.97 & 13.37 & 13.26 & ... & ... & ...  \\
 2 &  7  & 19.03 & 17.02  & 16.19 & 14.96 & 14.39 & 14.08 & ... & ... & ...  \\
 2 &  10  & 17.28 & 15.69  & 15.37 & 14.42 & 13.90 & 13.76 & ... & ... & ...  \\  
 2 &  11  & 16.55 & 14.59  & 13.76 & 13.45 & 12.77 & 12.67 & ... & ... & ...  \\  
 2 &  12  & ... & ...  & ... & 11.83 & 11.26 & 11.19 & 10.95 & 10.97 & 9.86  \\  
 2 &  13  & ... & ... & ... & 14.52 & 13.34 & 12.83 & ... & ... & ...  \\  
 2 &  14  & ... & ...  & ... & 14.50 & 13.58 & 13.33 & ... & ... & ...  \\  
 2 &  19  & 19.31 & 16.79  & 15.77 & 14.14 & 12.92 & 12.38 & 12.24 & 12.38 & 8.62  \\  
 2 &  20  & 21.07 & 18.88  & 18.12 & 15.57 & 14.26 & 13.46 & ... & ... & ...  \\   
 2 &  21  & ... & 14.30  & 16.11 & ... & ... & ... & ... & ... & ...  \\   
  \hline
 1 &   27 & ... & ... & ... & 13.26 & 12.70 & 12.53 & ... & ... & ... \\
 1 &    31 & 17.45 & 14.94 & 14.01 & 12.99 & 12.44 & 12.13 & ... & ... & ... \\
 2 & 22  & 17.17 & 15.14  & 14.04 & 13.05 & 12.41 & 12.20 & ... & ... & ...  \\ 
 \label{17MW}
\enddata
\tablecomments{``..." indicates that the counterpart was not seen at the corresponding wavelength. Sources at the bottom of the table, below the horizontal line, correspond to sources added in by hand (see Section \ref{data}).}
\tablenotetext{\rm a}{Magnitudes taken from the USNO-B optical catalog.}
\tablenotetext{\rm b}{Magnitudes taken from the Two-Micron All Sky Survey (2MASS).}
\tablenotetext{\rm c}{Magnitudes taken from the Wide-Field Infrared Survey Explorer (WISE).}
\tablenotetext{\rm d}{The MW counterpart to this source lies outside of the X-ray source's positional error circle. See Field 1 Source 6 in Section \ref{result} for more details. }
\tablenotetext{\rm e}{Marks sources that have more than one optical source within their positional uncertainty}
\end{deluxetable*}

\clearpage

\section{Detected {\sl Chandra} Sources }

Out of the 56 sources listed in Table \ref{tab1} only five had enough counts ($\geq100$) for meaningful spectral fitting. The count rate spectra of these sources have been created using the  {\tt specextract}   CIAO tool called by the {\tt srcflux} CIAO script. This script creates circles with differing\footnote{The CXO PSF becomes wider with increasing off-axis angle} radii that enclose 90\% of the PSF at 1 keV at a given location on the chip. The background is also automatically extracted by the {\tt srcflux} script from adjacent annular regions. We fit each of the five spectra with a blackbody (BB), power-law (PL), and APEC \citep{2001ApJ...556L..91S} models and report the results of the best-fit models in Table \ref{sparam}. We show the best-fit spectra in Figure \ref{spectra}. Due to the low number of counts for sources 18 and 25 in Field 1, their spectra were left unbinned and fit using C-statistics.\footnote{\url{https://heasarc.gsfc.nasa.gov/xanadu/xspec/manual/XSappendixStatistics.html}} 

\label{result}
\subsection{Brighter Sources in Fields 1 and 2}

\subsubsection{Field 2 Source 1 (CXOU J161423.4--505738)}

Source 1 in Field 2 has 663 background subtracted counts, which makes it the brightest source in either of the two fields. The spectrum for this source was extracted from an $r=10''$ circle,
and the background was taken from an $10''<r<50''$ annulus around the source. The spectrum is well fit ($\chi^2_{\nu}=$1.1 for $\nu=58$ degrees of freedom) by a PL model with a photon index $\Gamma=0.2\pm0.3$ and intervening hydrogen absorption column $N_{H}=3.6^{+0.9}_{-0.8}\times10^{22}$ cm$^{-2}$ (see Figure \ref{spectra}). The total Galactic HI column density in the direction of this source is $N_{H}=2.1\times10^{22}$ cm$^{-2}$ \citep{1990ARA&A..28..215D}, suggesting that this source is at a distance of at least a few kpc. In fact, the fitted $N_{H}$ is almost identical to the $N_{H}$ found by K+09 in their spectral fits for PSR J1617, implying that they may lie at similar distances. If we assume that this source is at a distance of 5 kpc, it would then have an X-ray luminosity $L_{\rm 0.5-7keV}=1.4\times10^{33} d^2_{5}$ erg s$^{-1}$. The nearest object to this source at other wavelengths is 5\farcs4 away and therefore is unlikely to be related to the X-ray source.

Despite the very hard spectrum in the 0.5-7.0 keV band, there is no {\sl INTEGRAL/IBIS} counterpart for this source down to a limiting flux of 3.7$\times10^{-12}$ erg s$^{-1}$ cm$^{-2}$. Using the online WebPIMMS tool\footnote{\url{https://heasarc.gsfc.nasa.gov/cgi-bin/Tools/w3pimms/w3pimms.pl}}, we have extrapolated our {\sl Chandra} fit in the 0.5-7.0 keV band to the {\sl INTEGRAL} 17-60 keV band and found that the expected flux should be a factor of $\sim$ 4-11 higher than the {\sl INTEGRAL} catalog detection limit for a photon index between  0.5 and 0.0, respectively. This suggests that the spectrum has a break (or a cut-off) at an energy that lies within the hard X-ray band (10-60 keV) or that the source is variable.

Periodicities have been found in sources with similarly hard X-ray spectra (e.g., \citeauthor{2010A&A...523A..50F} \citeyear{2010A&A...523A..50F}), so we have carried out a periodicity search for this source. A Z$^{2}_1$ test \citep{1983A&A...128..245B} was used to search for a periodic signal. First, the arrival times of the photons were corrected to the solar system barycenter using the CIAO tool {\tt axbary}. We then searched $\sim$10$^4$ equally spaced frequencies between $\nu\approx$0.16--3.4$\times10^{-5}$ Hz. The largest Z$^{2}_1= 21.3$ ($77\%$ signal detection confidence) was found at a period of 55.4 s, implying that there is no hard evidence of a periodic signal in the period range of 6.5 s--8.2 ks. We have also found no signs of variability in the light curve of this source using a number of different sized bins.

It is still possible that this source is either a low mass X-ray binary (LMXB) or a cataclysmic variable (CV) with a companion that is too absorbed and too distant to be detected at other wavelengths. The hardness of the spectrum makes this source less likely to be a LMXB. However, CVs with such hard X-ray spectra, a spectral break in the hard X-ray band, and counterparts too faint to be detected by the MW surveys used here have been discovered (see e.g., the intermediate polar IGR J18293--1213; \citeauthor{2016MNRAS.461..304C} \citeyear{2016MNRAS.461..304C}). This source is largely offset ($\sim 19'$) from the center of the TeV source making it unlikely to be related to the TeV emission. Furthermore, although there have been claims of TeV emission from CVs (see \citeauthor{2011MNRAS.411.1701B}  \citeyear{2011MNRAS.411.1701B} and references therein), they are not known to be extended TeV sources. This source's spectrum is also too hard for an isolated pulsar, which could otherwise power a relic PWN. Due to the lack of MW information, our MW classification tool was unable to confidently classify this source. Future observations in the hard X-ray band and deeper NIR/IR observations could shed more light on the this source's nature.

\vspace{-3mm}
\subsection{Field 1 Source 6 (CXOU J161723.7--505152)}

The spectrum for this source was extracted from an $r=$ 7$''$ circle centered on the source. The background was extracted from a $7''<r<35''$ annulus. There are 184 net counts in the $0.5-7$ keV band. This source has no optical, NIR, or IR counterpart within its 2$\sigma$ positional error radius (0$\farcs$8). However, there is a star (HD 146183, $B=10.38$) that is $1\farcs7$ (based on the Gaia position) offset from the X-ray position\footnote{The star's USNO--B coordinates are $1\farcs3$ offset from the X-ray source position}. Despite the fact that HD 146183 is just outside of the 2$\sigma$ X-ray positional error circle,  Source 6's soft X-ray spectrum suggests that the emission can be associated with the corona of the bright, nearby star.  Therefore, we consider the star as the MW counterpart of Source 6. The MW magnitudes can be seen in Table \ref{17MW} and are included when classifying this source with our pipeline. HD 146183 has a rather large proper motion listed in the Gaia catalog $\mu_{\alpha}=(-20.521)$ mas yr$^{-1}$, $\mu_{\delta}$=(-27.979) mas yr$^{-1}$ and has a spectral type F0IV (\citeauthor{2016A&A...595A...4L} \citeyear{2016A&A...595A...4L}; \citeauthor{2000A&A...355L..27H} \citeyear{2000A&A...355L..27H}; \citeauthor{1978mcts.book.....H} \citeyear{1978mcts.book.....H}). The parallax distance to this star is 300 pc \citep{2016A&A...595A...4L}.

Of the three models, the spectrum is best fit by the APEC model, but there is an excess of high energy photons. A better fit can be obtained by adding a second, higher temperature component. However, due to the small number of counts, the uncertainties in the second component's parameters are too large to derive any meaningful constraints, so this fit is omitted. The best-fit parameters are given in Table \ref{sparam}.  We found no evidence of variability in the light curve of this source. Our classification pipeline also classifies this source as a star with a high confidence (99\%). The parallax distance to HD 146183 implies an X-ray luminosity L$_{\rm 0.5-7kev}=5.7\times10^{29}$ erg s$^{-1}$, which is consistent with X-ray luminosity of a coronally active star. This makes Source 6 an unlikely counterpart to the TeV source. 

\vspace{-3mm}
\subsection{Field 1 Source 13 (CXOU J161551.0--504925)}

Source 13's spectrum was extracted from an $r=7''$ circle, while the background was taken from a $7''<r<34''$ annulus. This source has 388 net counts and is the second brightest source detected in either of the {\sl Chandra} fields. It should be noted that Source 13 is the closest source with $>$100 counts to the center of the TeV emission ($\sim 7'$ offset). There is a very bright ($B$=14.42, $J$=9.88, $W1$=8.80) counterpart detected in USNO-B, 2MASS, and {\sl WISE}, which is offset from the X-ray position by $\sim0\farcs5$. It is also found in The Second-Generation Guide Star Catalog \citep{2008AJ....136..735L}. This object has a measured proper motion in the UCAC4 catalog of $\mu_{\alpha}\cos{\delta}=(-7.7\pm1.8)$ mas yr$^{-1}$, $\mu_{\delta}=(-8.5\pm1.8)$ mas yr$^{-1}$ \citep{2012yCat.1322....0Z}, ruling out the possibility that it is an AGN. It is possible, although unlikely, that the optical/IR source is accidentally projected at the X-ray source position. However, the X-ray hardness ratios\footnote{The hardness ratios are defined as HR2=$(F_{\rm 1.2-2 keV}-F_{\rm 0.2-1.2 keV})/(F_{\rm 1.2-2 keV}+F_{\rm 0.2-1.2 keV})$ and HR4=$(F_{\rm 2-7 keV}-F_{\rm 0.5-2 keV})/(F_{\rm 2-7 keV}+F_{\rm 0.5-2 keV})$.} also rule out the possibility that this source is an obscured AGN undetected in the optical/NIR (see Figure \ref{src118}). The X-ray spectrum is well fit by either an absorbed APEC or PL model. The APEC model produces a temperature of $kT=4.1^{+1.1}_{-1.0}$ keV and absorption column $N_{H}=(2^{+0.2}_{-0.1})\times10^{21}$ cm$^{-2}$, while the best-fit absorbed PL model has a photon index $\Gamma=2.3\pm0.3$ and absorption column $N_{H}=(4\pm2)\times10^{21}$ cm$^{-2}$ (see Figure \ref{spectra} for the spectrum and best-fit PL model). 

The large amount of MW data provided for this source allowed us to fit for a photometric spectral classification. We used the SEDfitter\footnote{https://github.com/astrofrog/sedfitter} \citep{2007ApJS..169..328R} code to fit the photometry of this source to stellar models from \cite{2004astro.ph..5087C}. The best fit models for this source are those of M0V--M3V M-dwarf stars. The X-ray to $B$ band luminosity ratio $\log(L_{\rm 0.5-2 keV}/L_{B})=-3.1$ is also consistent with a late type star. Our classification pipeline also classified this source as a star with a confidence of 72\% (the Guide Star Catalog also classifies this source as a star). The distance to this source, using the source's V band magnitude V=13.04 \citep{2012yCat.1322....0Z}, is 74--29 pc giving and X-ray luminosity $L_{\rm 0.5-7keV}\sim8-1\times10^{28}$ erg s$^{-1}$ for M0V and M3V spectral types, respectively. This X-ray luminosity is within the typical range for M-type stars (see e.g., \citeauthor{2004A&ARv..12...71G} \citeyear{2004A&ARv..12...71G} and \citeauthor{2005ApJS..160..390P} \citeyear{2005ApJS..160..390P}). We have also searched for variability in the light curve of this source but none was found. The classification as a late-type dwarf star makes this source an unlikely counterpart to the TeV source.

\subsection{Field 1 Source 18 (CXOU J161643.5--504604)}

The spectrum for this source was extracted from a circle surrounding the source with a radius of $r= 5''$. The background was extracted from a $5''<r<23''$ annulus. In total, this source has 116 net counts. The best fit model for this source is an absorbed PL with $\Gamma=1.8\pm0.5$ and $N_{H}=(5\pm4)\times10^{21}$ cm$^{-2}$. The nearest optical/NIR source is 4.8$''$ away, making it an unlikely MW counterpart.

\begin{figure*}
\begin{center}
\includegraphics[scale=0.6,trim=0 0 0 0,angle=270]{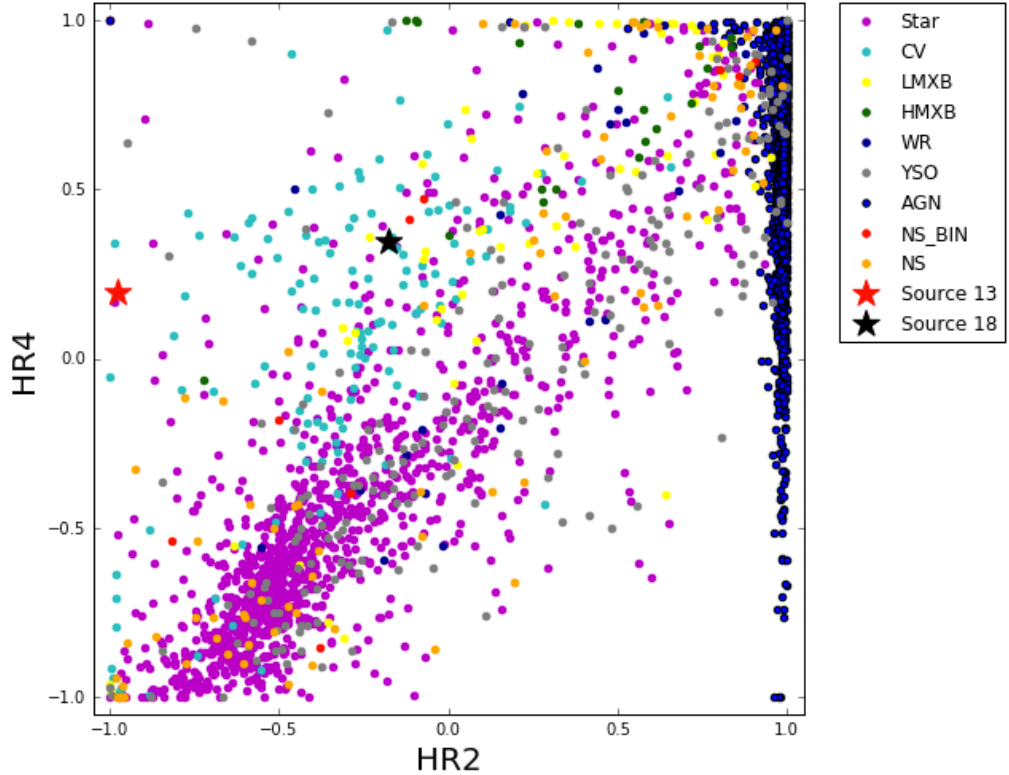}
\caption{Positions of Sources 13 and 18 from Field 1 on the hardness-ratio diagram plotted over our training dataset. The AGN fluxes were corrected for the extinction in the direction of HESS J1616 before calculating the hardness ratios. Both Source 13 and 18 lie far away from AGN in this parameter space. No extinction correction was attempted for the training dataset sources from other classes because, in many cases, their distances are unknown. The classes are defined in the appendix of \cite{2016ApJ...816...52H}.}
\label{src118}
\end{center}
\end{figure*}

\begin{figure*}
\begin{center}
\includegraphics[scale=0.30,trim=0 0 0 0]{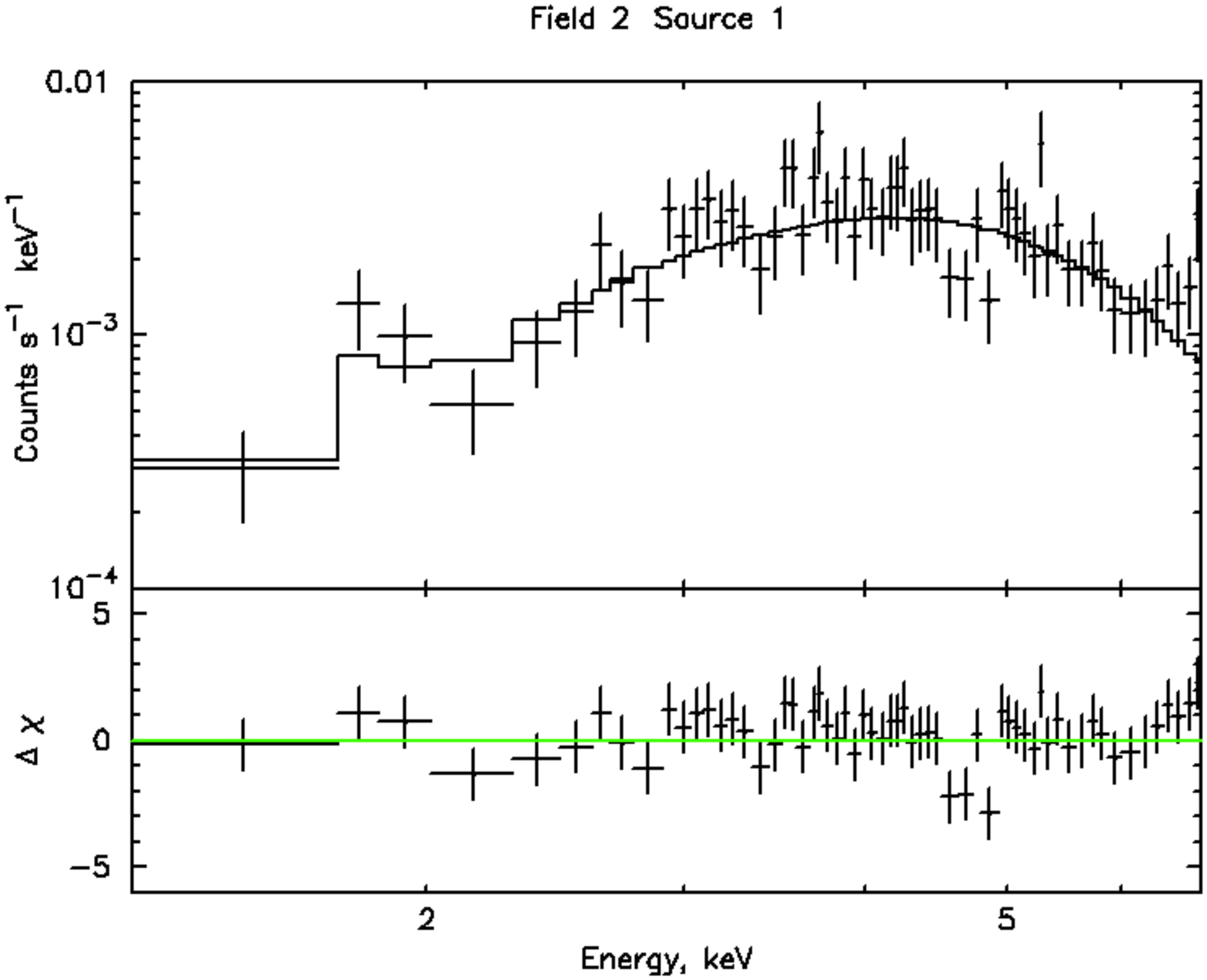}
\includegraphics[scale=0.309,trim=0 2 0 0]{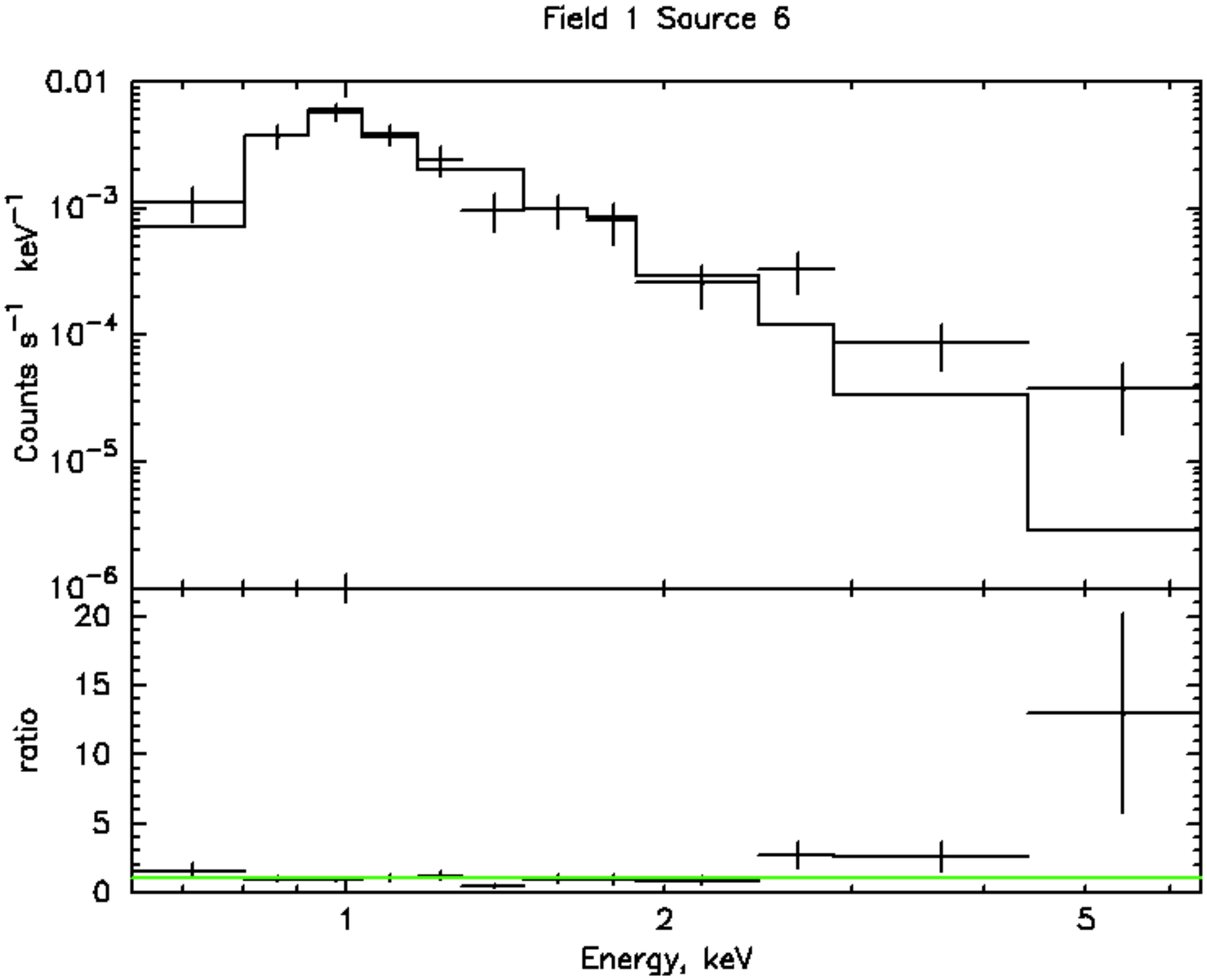}
\includegraphics[scale=0.303,trim=0 10 0 0]{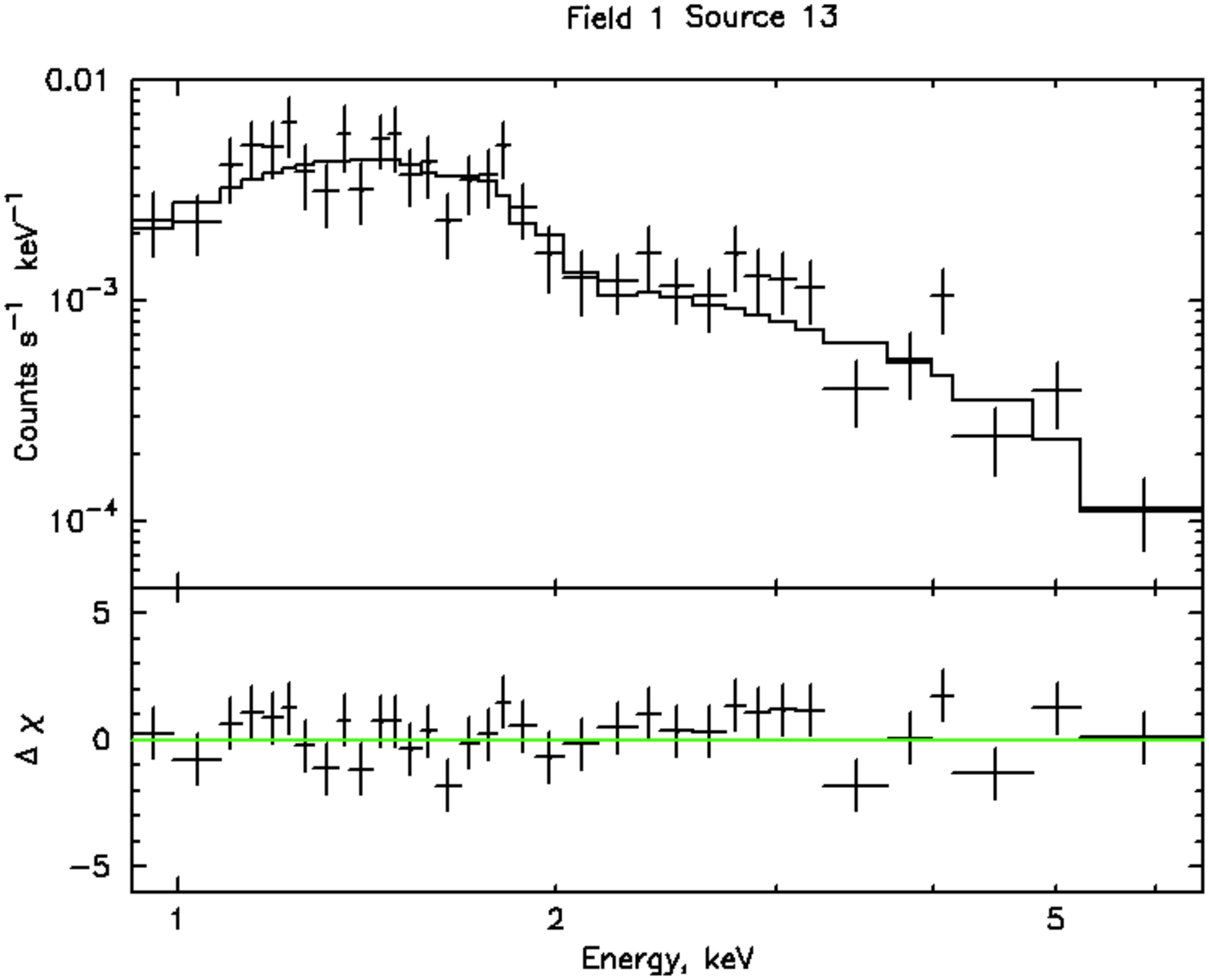}
\includegraphics[scale=0.3045,trim=0 10 0 0]{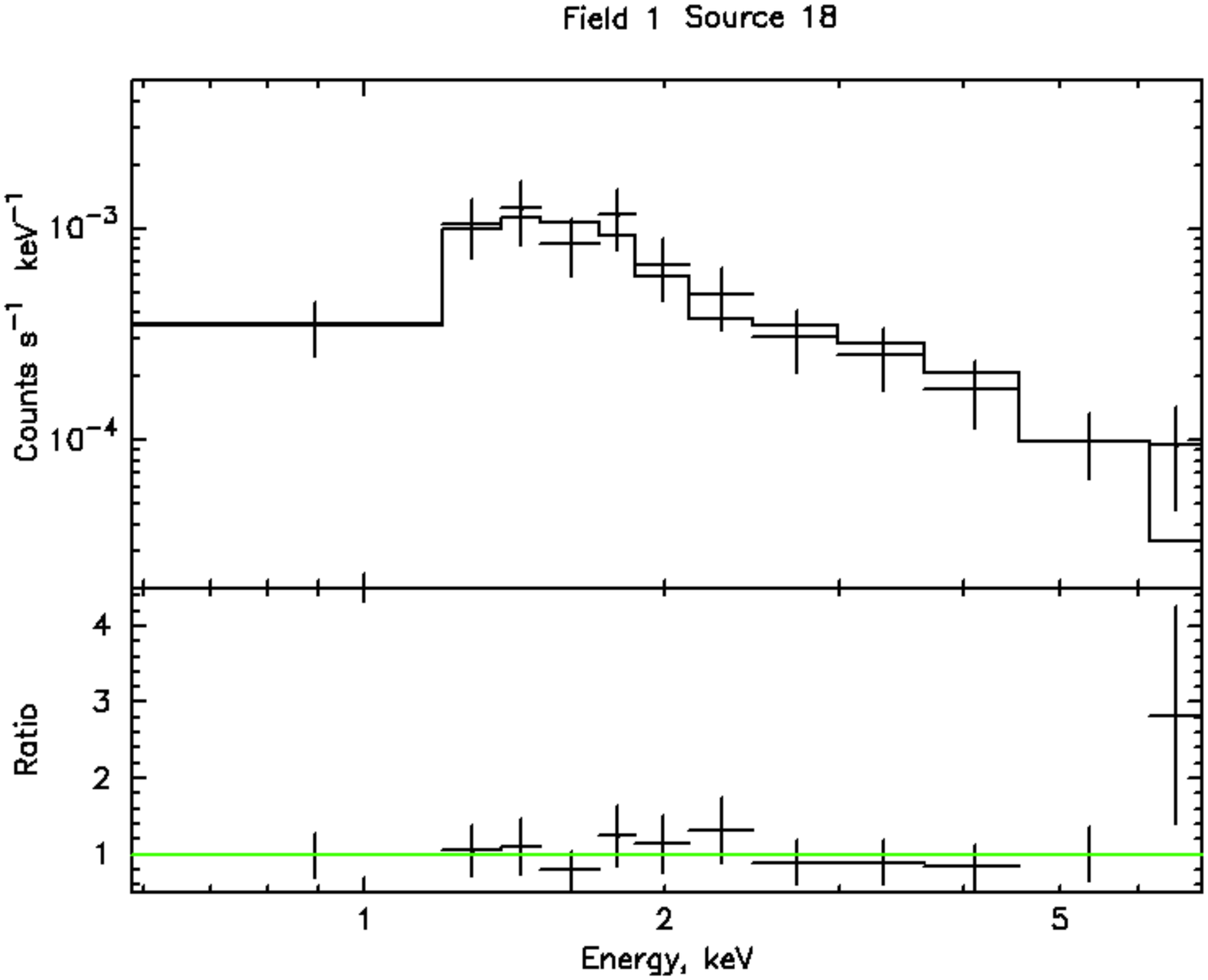}
\includegraphics[scale=0.31,trim=0 0 0 0]{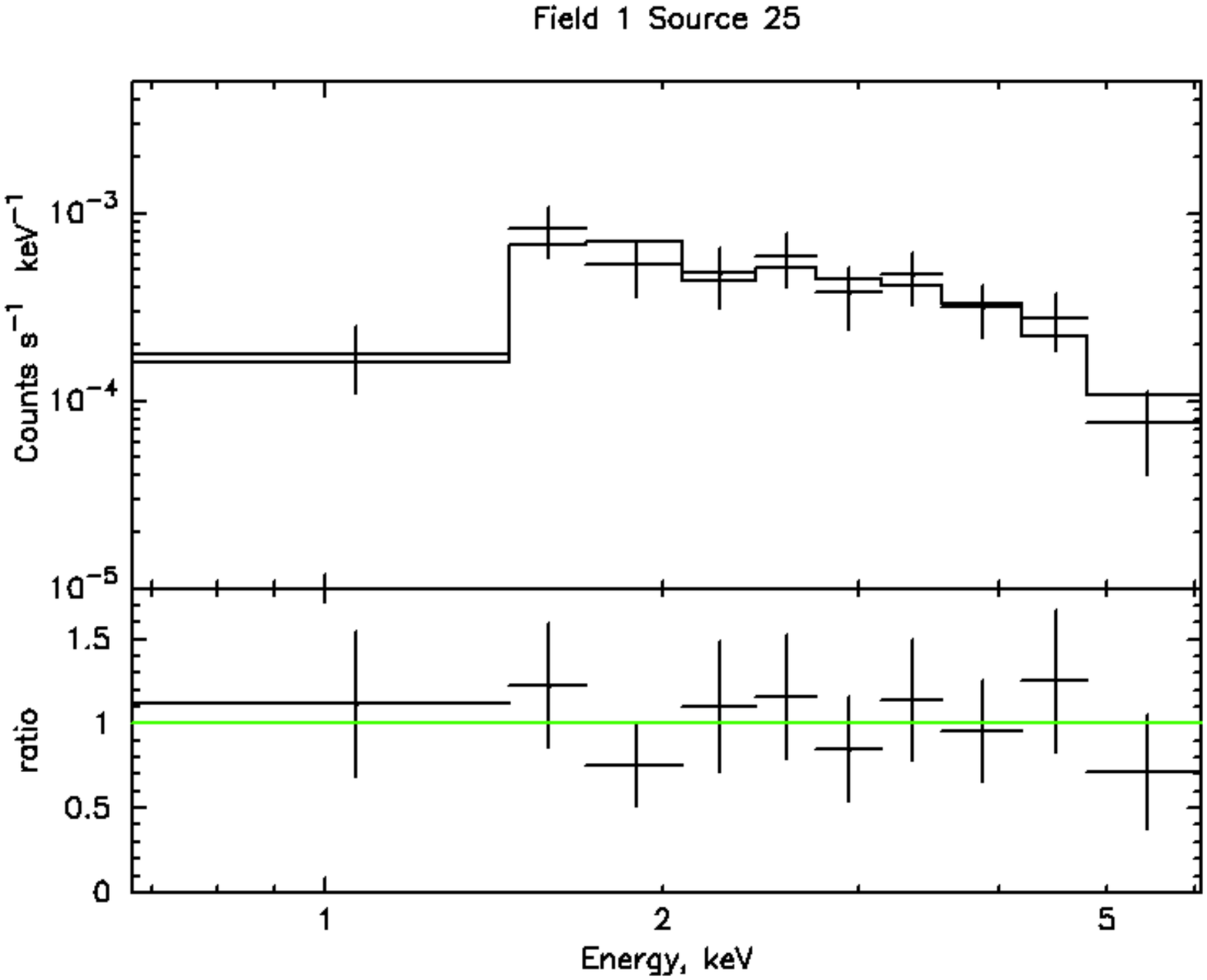}
\caption{Spectral fits for 5 sources with $\geq$ 100 counts. Source 1 and Source 13 were fit using chi-squared statistics with 10 counts per bin. The other three spectra were left unbinned and fit using Cash statistics (C-stat) and the binned spectra are only shown for visualization.}
\label{spectra}
\end{center}
\end{figure*}

The absorption column density ($\sim$ 1/4 of the total Galactic value) suggests that this source is fairly nearby, assuming that the chosen spectral model is meaningful. The source spectrum and lack of a counterpart is consistent with an isolated pulsar interpretation. Therefore, it could be a relatively nearby unknown pulsar with a relic PWN. The relative proximity would relax the demands for the TeV luminosity and large physical size of the TeV source. If we assume that this source and HESS J1616 are associated we get, then the TeV to X-ray luminosity ratio $L_{\rm 1-10TeV}/L_{\rm 0.5-8keV}=340$. This  is consistent with known pulsars with relic PWN and associated TeV emission (see Figure 4 in \citeauthor{2013arXiv1305.2552K} \citeyear{2013arXiv1305.2552K}). However, no hints of extended X-ray emission around this source are detected. We have plotted the hardness ratios of Source 18 on top of the sources in our training dataset (see Figure \ref{src118}). The AGN fluxes used in the calculation of the hardness ratios in this plot have been corrected for the extinction in the direction of HESS J1616. Source 18 lies within a parameter space far away from the AGN, implying that is unlikely to be an AGN. Due to the lack of MW counterparts our pipeline was unable to confidently classify this source. More X-ray data and deeper MW limits are needed to identify the nature of this object.

\vspace{-3mm}
\subsection{Field 1 Source 25 (CXOU J161703.6--504705)}

The spectrum for this source was extracted from a circle surrounding the source with a radius of $r=5''$. The background was extracted from a $5''<r<27''$ annulus surrounding the source. This source only has 100 net counts and was fit using C-statstics. The spectrum can be equally well fit by all three models. The best fit parameters for each model can be seen in Table \ref{sparam}. This source has a likely NIR counterpart in the 2MASS Extended Catalog (2MASX; \citeauthor{2006AJ....131.1163S} \citeyear{2006AJ....131.1163S}) and is detected at 5 GHz in radio with a flux S($\nu$)$=$3.3 Jy (\citeauthor{1987A&A...171..261C} \citeyear{1987A&A...171..261C}; see Figure \ref{MWfig} for an 843 MHz radio image). The X-ray source (see Figure \ref{cxofig}), which is possibly surrounded by some diffuse X-ray emission, is embedded within the extended NIR/IR emission ($\sim70''$ half-light radius in the $J$ band) seen in the {\sl Spitzer} image (see Figure \ref{MWfig}). The likely MW counterpart is a known young, massive stellar cluster with two components (the first component being brighter by a factor of $\sim$10 in the NIR) mostly made up of zero-age main-sequence O8 and B0.5 type stars, respectively \citep{2004AJ....127.2817R}. Our automated classification pipeline  supports the association of the X-ray source with the star forming region and has classified this source as a YSO.

There are several O spectral type stars located in this cluster, suggesting that the cluster is relatively young (a few Myr). This cluster is embedded in an HII region, which has an upper limit on its age of $<$10 Myr \citep{2016A&A...588A..63C}. The X-ray luminosity of this source (using the distance of 3.7 kpc assumed by \citeauthor{2004AJ....127.2817R} \citeyear{2004AJ....127.2817R}) is $L_{\rm 0.5-7keV}\approx6\times10^{31}$ erg s$^{-1}$, which is typical an O type star or small group of O-type stars (see e.g., \citeauthor{2011ApJS..194....7N} \citeyear{2011ApJS..194....7N}). It is possible that a young, energetic NS or a BH could have been produced in this cluster, but we find no evidence of it. Another possibility is that the TeV emission could be produced by massive colliding wind binaries (see e.g., \citeauthor{2006ApJ...644.1118R} \citeyear{2006ApJ...644.1118R}). However, the TeV luminosity of HESS J1616 is too large ($\sim 3\times10^{34}$ erg s$^{-1}$ at 3.7 kpc) to be consistent with this scenario, assuming a similar particle injection spectrum and model as in \cite{2016A&A...591A.139D}. Also, the size of the TeV source is much larger than the cluster size. Therefore, we conclude that the most likely cause of the X-ray emission is that it is coming from a group of YSOs or  O/B$-$type stars and their winds. Even if some of the TeV emission could be attributed to the shocks driven by winds of massive stars, it is likely to be a relatively small fraction of the TeV emission from HESS J1616.

\begin{deluxetable*}{ccccccc}
\tablecaption{Best-fit spectral parameters for sources with $\geq$ 100 counts.}
\tablewidth{0pt}
\tablehead{
\colhead{Field} & \colhead{Source} & \colhead{Model\tablenotemark{a}} & \colhead{Norm\tablenotemark{b}} & \colhead{$N_{H}$ cm$^{-2}$} & \colhead{$\Gamma$/kT} & \colhead{$\chi^2$ or C-stat (C)/d.o.f.}} 
\startdata
& & & $\times$10$^{-5}$ & $\times$10$^{22}$ & $.../{\rm keV}$ & \\
\hline
2 & 1 & PL & 2.0$^{+1.4}_{-0.8}$ & 3.6$^{+0.9}_{-0.8}$ & $\Gamma=0.2\pm0.3$ & 64.16/58 \\
1 & 6 & AP & 2.4$^{+0.6}_{-0.5}$ & 0.2$+0.1$ & kT$	=0.96^{+0.06}_{-0.07}$ & 114.01/91 (C)\\
1 & 13 & PL & 4.7$^{+1.6}_{-1.2}$ & 0.4$\pm0.2$ & $\Gamma=2.3\pm0.3$ & 32.34/32\\
1 & 13 & AP & 8.1$^{+1.4}_{-0.9}$ & 0.2$^{+0.2}_{-0.1}$ & kT$=4.1^{+1.1}_{-1.0}$ & 32.99/32\\
1 & 18 & PL & 1.0$^{+0.9}_{-0.4}$ & 0.5$\pm0.4$ & $\Gamma=1.8\pm0.5$ & 89.96/90 (C)\\
1 & 25 & BB & 570$^{+560}_{-270}$ & 0.6$^{+0.5}_{-0.4}$ & kT$=0.9\pm0.2$ & 62.76/94 (C)\\
1 & 25 & PL & 3.0$^{+5.1}_{-1.8}$ & 2.0$^{+0.9}_{-0.8}$ & $\Gamma=2.3\pm0.7$ & 62.94/94 (C)\\
1 & 25 & AP & 5.5$^{+2.7}_{-1.7}$ & 1.8$^{+0.7}_{-0.6}$ & kT$=3.7^{+5.3}_{-1.4}$ & 62.28/94 (C)\\
\label{sparam}
\enddata
\tablenotetext{}{All uncertainties are at 68\% confidence for each single interesting parameter. Sources whose spectra are well fit by multiple models have all acceptable fits listed in the Table. }
\tablenotetext{a}{The models are abbreviated as follows, power-law (PL), APEC (AP), and blackbody (BB). }
\tablenotetext{b}{The normalizations are defined differently for each model. The BB normalization is defined as $N=R_{\rm km}^2/D_{\rm 10kpc}^2$, where R is the source radius in units of km and D is the distance to the source scaled to 10 kpc. The PL normalization $N$ is given in units of photons keV$^{-1}$ cm$^{-2}$ s$^{-1}$ at 1 keV. The normalization of the AP model is defined by the integral $N=10^{-14}/(4\pi[D_A(1+z)^2])\int n_{\rm e} n_{\rm H} dV$, with D$_A$ defined as the distance to the source in cm, n$_{\rm e}$ is the electron density in cm$^{-3}$ and n$_{\rm H}$ is the hydrogen density in cm$^{-3}$. }
\end{deluxetable*}

\subsection{Remaining Sources}
Our MW machine-learning pipeline was able to confidently (i.e., $\geq70\%$ confidence) classify 32 of the remaining 51 sources that were not bright enough to fit a constraining spectrum. The majority (16) of these sources were classified as stars while the others were classified as either AGN (12) or YSOs (4). 
 All but one of the sources classified as an AGN have no MW counterparts, and none have fluxes in the ultra-soft or soft {\sl CXO} bands, implying heavy absorption. Therefore these sources are likely AGN, but we cannot entirely rule out the possibility of an obscured X-ray binary in quiescence at a large distance in our Galaxy. AGN and X-ray binaries can be TeV emitters, but these sources are typically point-like and cannot be responsible for the extended nature of the TeV source if it is a single source. Stars and YSOs are not known to produce TeV emission, so we can rule out these sources.

The closest source to ``Source X'' discussed in K+09 is Source 29 in Field 1. This source is surrounded by a number of faint point sources and possibly faint extended emission. The X-ray source is coincident with a star forming region that can be seen in the {\sl Spitzer} image in Figure 1. Therefore, we rule out ``Source X'' as a possible counterpart to the TeV source.

\section{ Non-detection of PSR J1614--5048}
\label{J1614}

\begin{figure*}
\begin{center}
\includegraphics[scale=0.438,trim=0 50 0 0]{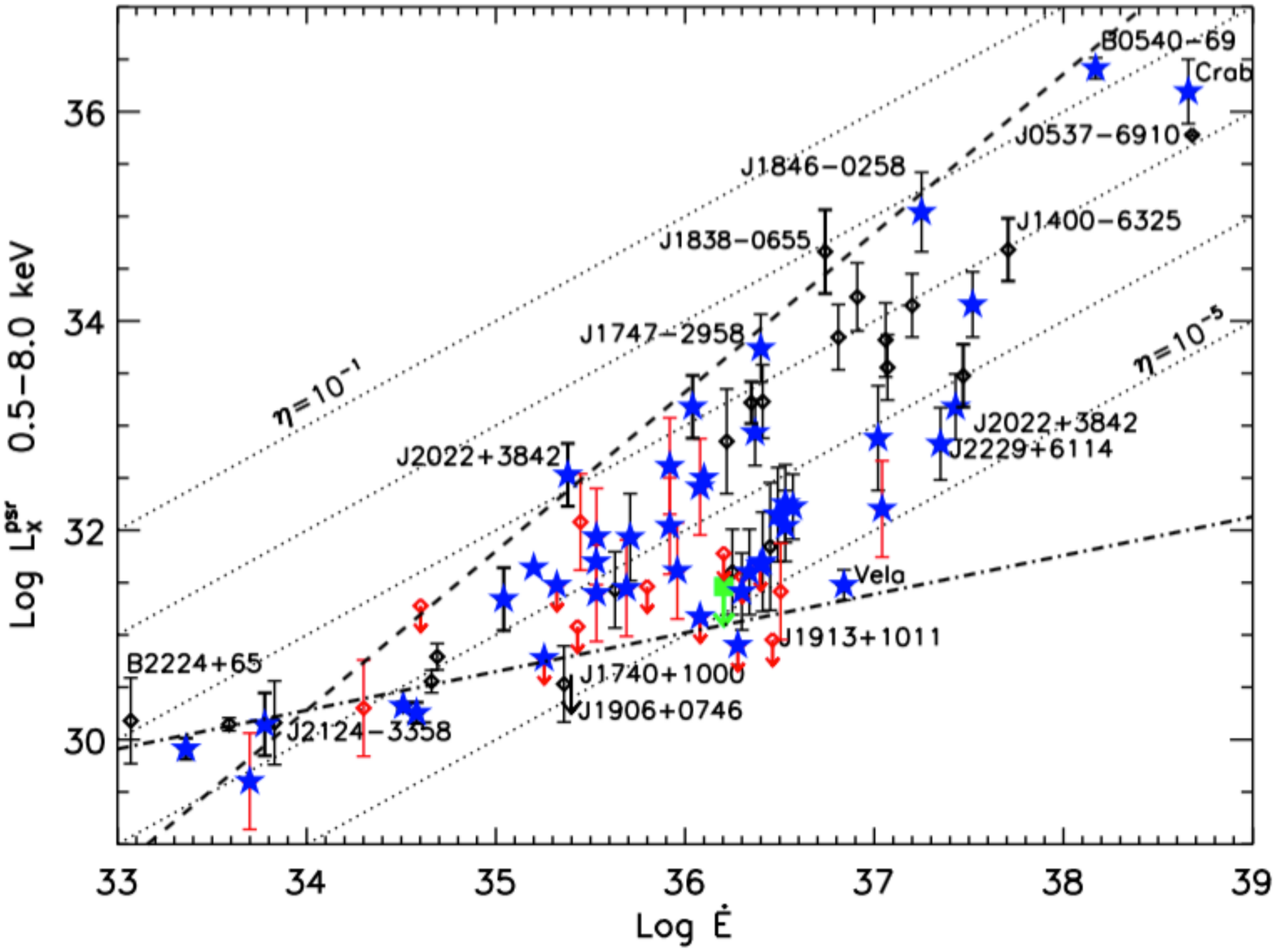}
\includegraphics[width=11.8cm, height=10cm,trim=14 50 0 0]{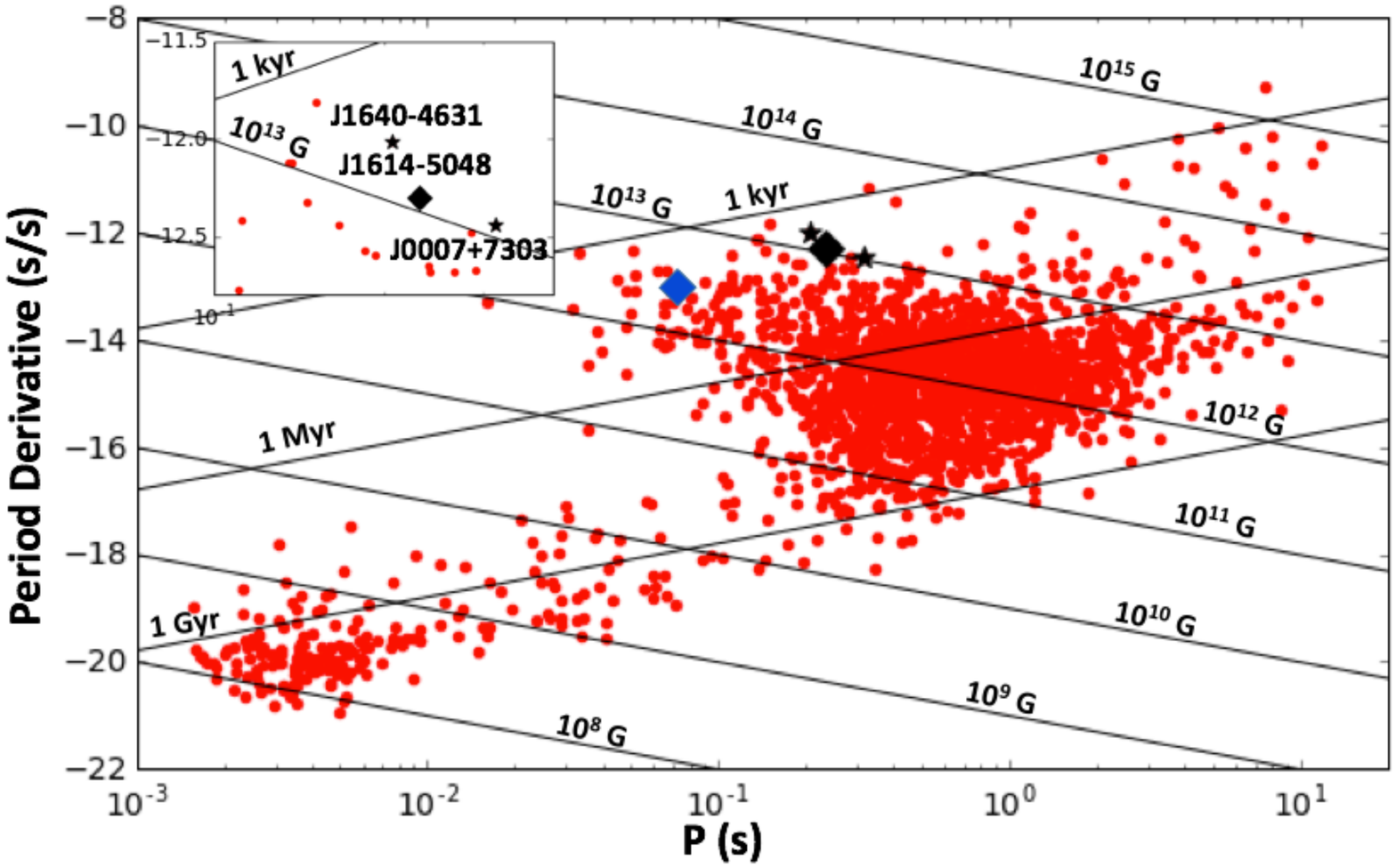}
\caption{{\sl Top:} X-ray luminosity versus $\dot{E}$ for a sample of pulsars shown in Figure 2 of \cite{2012ApJS..201...37K}. The dashed and dash-dotted lines correspond to $\log{L_{X}^{\rm psr}}=1.51\log{\dot{E}}-21.4$ and $\log{L_{X}^{\rm psr}}=0.38\log{\dot{E}}-17.7$, respectively. The constant efficiency ($\eta=L/\dot{E}$) lines are shown by the dotted lines, and the downward arrows show the 90\% confidence upper limits. The blue stars mark $\gamma$-ray pulsars \cite{2012ApJS..201...37K}. The upper limit on the luminosity of J1614--5048 (see Section \ref{J1614}) is shown by the green square with a downward arrow. {\sl Bottom:} $P$--$\dot{P}$ diagram showing PSR J1614--5048 as a black diamond and PSR J1617--5055 as a blue diamond. The high-B pulsars compared to PSR J1614-5048 in Section \ref{J1614} (PSR J1640--4631 and PSR J0007+7303) are shown as black stars. The inset on the top left is zoomed in on the region including these pulsars. Pulsar parameters were taken from the ATNF catalog \citep{2005AJ....129.1993M}.}
\label{xefffig}
\end{center}
\end{figure*}

PSR J1614 was imaged $\sim 0\farcm8$ away from the optical axis in Field 2. We find 5 photons with energies between 0.5 and 8 keV\footnote{This energy range was used instead of 0.5-7 keV for the purpose of comparison with other pulsars.} within the source region defined as an $r=2\farcs1$ circle centered on the interferometric radio position, $\alpha=16^{\rm h}14^{\rm m}11{\fs}6(8)$, $\delta=-50^{\circ}48\farcm01\farcs9(1)$ \citep{1999MNRAS.308..609S}. The local background estimated from the annular, $2\farcs1<r<35''$, region centered on J1614 is 0.085$\pm0.005$ counts arcsec$^{-2}$. On average, $\sim$one count within the source region is expected to be due to the background implying a detection with only $\sim 2\sigma$ source significance. Therefore, we calculate a 90\% confidence upper limit of 0.15 counts ks$^{-1}$ on the count rate from PSR J1614 and its possible compact PWN using Table 1 in \cite{1986ApJ...303..336G}. To calculate the upper limit on the unabsorbed flux in the 0.5--8 keV band, we used PIMMS and assumed an absorbed PL model with $\Gamma=2$ and $N_H=1.75\times10^{22}$ cm$^{-2}$ derived from the pulsar's DM=582.8 cm$^{-3}$ pc according to equation 2 from \cite{2013ApJ...768...64H}. The estimated upper limit of $5.0\times10^{-15}$ erg cm$^{-2}$ s$^{-1}$ corresponds to an X-ray luminosity $L_{\rm 0.5-8 keV}=2.9\times10^{31}$ erg s$^{-1}$ at an assumed distance $d=7$ kpc. This makes PSR J1614 one of the least efficient X-ray pulsars known, with an upper limit on its X-ray efficiency of $\eta_{\rm 0.5-8 keV}\equiv L_{\rm 0.5-8 keV}/\dot{E}\sim2\times10^{-5}$ (see Figure \ref{xefffig}, top panel).

PSR J1614 is not detected at GeV energies, which is similar to some other pulsars with small X-ray efficiencies (e.g., PSR J1906+0746, PSR J1913+1011). One exception is the Vela pulsar \citep{2001ApJ...552L.129P}, which has a small X-ray efficiency and is detected at GeV energies. However, Vela is much closer ($\sim 300$ pc; \citeauthor{2003ApJ...596.1137D} \citeyear{2003ApJ...596.1137D}) than the aforementioned pulsars. To understand how PSR J1614 compares to other similar pulsars, we have looked at the nearest pulsars on the $P$--$\dot{P}$ diagram\footnote{\url{http://www.atnf.csiro.au/people/pulsar/psrcat/}} (see Figure \ref{xefffig}, bottom panel) with comparably high magnetic fields (i.e., $\gtrsim10^{13}$ G). Unlike PSR J1614, neither of the two neighboring pulsars in the $P$--$\dot{P}$ diagram (i.e., J1640--4631 and J0007+7303) are detected at radio wavelengths, which could be due to a low intrinsic radio efficiency or a misalignment of their radio beams with the line of sight. 

Interestingly, PSR J1640--4631 has been associated with the TeV source HESS J1640--465 and the corresponding 1FHL source J1640.5--4634 \citep{2014ApJ...788..155G}. This pulsar has been detected by {\sl XMM-Newton},  pulsations were discovered by {\sl NUSTAR} \citep{2014ApJ...788..155G}, and a PWN was resolved with {\sl Chandra} \citep{2009ApJ...706.1269L}.  This pulsar is a factor of two younger than PSR J1614 (based on the spin-down ages) but it may be a factor of two more distant ($d=12$ kpc based on the DM; \citeauthor{1993ApJ...411..674T} \citeyear{1993ApJ...411..674T}). However, the younger age and slightly higher spin-down luminosity $\dot{E}=4.4\times10^{36}$ erg s$^{-1}$ (a factor of 2.8 larger than that of PSR J1614) can hardly explain the 100 times higher X-ray efficiency ($\eta_{\rm 0.5-8 keV}\approx2\times10^{-3}$). 

PSR J0007+7303 also has both an extended TeV counterpart detected by VERITAS (likely the PWN; \citeauthor{2013ApJ...764...38A} \citeyear{2013ApJ...764...38A}) and a {\sl Fermi-LAT} counterpart \citep{2012ApJ...744..146A} with pulsations detected, as well as extended emission, likely from the PWN. This pulsar and its wind nebula have also been detected by {\sl Chandra} \citep{2004ApJ...612..398H}. PSR J0007+7303 is a factor of two older than PSR J1614 with a characteristic age $\tau=13.9$ kyr, and it has a lower spin-down luminosity $\dot{E}=4.5\times10^{35}$ erg s$^{-1}$. It has a low X-ray efficiency $\eta_{\rm 0.5-10 keV}\approx10^{-5}$ comparable to that of PSR J1614.

Overall, it is interesting to note that several of the pulsars that are nearest to PSR J1614 on the $P$--$\dot{P}$ diagram have TeV sources associated with their PWN, implying that it may be possible for PSR J1614 to power a relic PWN that is responsible for the TeV emission. However, if we assume that PSR J1614 is responsible for the TeV emission, then the lower limit on the TeV to X-ray luminosity ratio of its undetected PWN, $L_{\rm 1-10TeV}/L_{\rm 0.5-8keV}>3400$, is larger than any other known pulsar with associated TeV emission (see Figure 4 in \citeauthor{2013arXiv1305.2552K} \citeyear{2013arXiv1305.2552K}). We can also compare the lower limit on the GeV to X-ray luminosity $L_{\rm 1-100 GeV}/L_{\rm 0.5-7 keV}\gtrsim11500$, which is larger than all but two pulsars (PSR J1836--5925 and PSR J2021+4026) in the {\sl Fermi} second pulsar catalog \citep{2013ApJS..208...17A}. The fact that PSR J1614 is an outlier in the TeV and GeV to X-ray flux ratios (assuming it is the counterpart to these sources), the lack of a PWN, and the large offset from the TeV source makes it an unlikely counterpart to HESS J1616.

\section{Conclusions and Outlook}
\label{conc}

TeV emission can be produced by a number of different source types, including blazars, SNRs, PWNe, and $\gamma$-ray binaries (see e.g., \citeauthor{2012APh....39...61H} \citeyear{2012APh....39...61H}). However, only PWNe and SNRs (or SNRs interacting with molecular clouds) are known to produce extended TeV emission. There are three SNRs detected at radio frequencies surrounding the center of HESS J1616 including Kes 32, RCW 103, and G332.0+0.2, but their rather large offsets ($\sim$ 14$'$, 15$'$, and 29$'$, respectively) and difference in morphology, compared to HESS J1616, make them unlikely counterparts to the TeV emission. Therefore, among the known types of TeV emitters, the only other candidates are relic PWNe, which can be dim in X-rays. The relic plerion scenario can also account for the offset between the known pulsars, within the extent of HESS J1616, and the TeV emission peak. However, until an X-ray or radio PWN with a preferential extension toward the TeV source is detected, this scenario will remain unconfirmed. Observations with the SKA may be able to detect synchrotron emission from such relic plerions.

The {\sl Fermi} source, 3FGL J1616.2--5054e, is coincident with the TeV emission. This source is extended and similar in size to HESS J1616. The spectra of the GeV and TeV sources smoothly connect to each other (see Figure 6.8 in \citeauthor{2014arXiv1401.6718L} \citeyear{2014arXiv1401.6718L}). The hard spectral index $\Gamma=1.74$ listed in the 2FHL catalog \citep{2016ApJS..222....5A} is typical of other PWNe detected by {\sl Fermi} \citep{2013ApJS..208...17A}.   
We have analyzed three {\sl Chandra} observations of the field containing HESS J1616 and searched among the 56 detected X-ray sources for a possible counterpart. Of these, 30 were classified with $\geq$70\% confidence, primarily as either AGN or non-degenerate stars. Stars are not known to produce TeV emission, while AGNs seen at TeV energies do not produce extended emission with the size of HESS J1616, so any of these sources can be ruled out as sole counterparts to HESS J1616. Five sources were bright enough ($\geq 100$ counts) to extract and fit their spectra. Of these, only Source 18 in Field 1 may be of interest because we cannot rule out an isolated pulsar interpretation. If this source is a nearby pulsar, it could host a relic PWN that could be responsible for the TeV emission. The TeV to X-ray luminosity ratio is compatible with other pulsars detected at TeV energies, but there are no hints of extended X-ray emission. Better quality X-ray data and deeper MW limits for a possible counterpart of this source are necessary to classify it.

We did not detect PSR J1614 and set an upper limit on its X-ray flux ($F_{\rm 0.5-8 keV}<5\times10^{-15}$ erg cm$^{-2}$ s$^{-1}$ at a 90\% confidence level). This makes PSR J1614 one of the least efficient X-ray pulsars known, with an upper limit on its X-ray efficiency $\eta_{\rm 0.5-8 keV}$ $\lesssim 2\times10^{-5}$. While, the lack of an X-ray PWN does not preclude the existence of a relic TeV PWN, it does not allow us to look for any PWN asymmetry, which could lend support to the association. Lastly, the large offset between PSR J1614 and HESS J1616 of $\sim 22'$ (45 pc  at  $d=$7 kpc) makes the association doubtful, so we consider PSR J1614 an unlikely counterpart to HESS J1616.

The only other known pulsar energetic enough to power a relic PWN in this region is PSR J1617. The offset from HESS J1616 is smaller for PSR J1617 ($10'$, is 19 pc at $d=6.5$ kpc). The faint extended X-ray PWN of PSR J1617 was detected by {\sl CXO}, but it does not extend in the direction toward the TeV source center making the association ambiguous (see K+09 for a discussion). If we assume that the PWN of PSR J1617 is responsible for the TeV emission, then the TeV to X-ray luminosity ratio is $L_{\rm 1-10 TeV}/L_{\rm 0.5-8 keV}\sim23$ and the GeV to X-ray luminosity ratio is $L_{\rm 1-100 GeV}/L_{\rm 0.5-8 keV}\sim91$. These ratios are consistent with other PWN observed at TeV energies and pulsars detected by {\sl Fermi} (\citeauthor{2013arXiv1305.2552K} \citeyear{2013arXiv1305.2552K}; \citeauthor{2013ApJS..208...17A} \citeyear{2013ApJS..208...17A}). Therefore, PSR J1617 remains a possible counterpart of the TeV source, given that the associations of the other X-ray sources in this field appear to be less likely.

Future observations with CTA and SKA will be able to shed more light onto this source's nature, in particular by testing wether it is a single extended or multiple blended source, and whether there is any spatially-dependent spectral structure or other morphological connection to PSR J1617.

\medskip\noindent{\bf Acknowledgments:}
Support for this work was provided by the National Aeronautics and Space Administration through Chandra Awards G05-16075X and AR3-14017X issued by the Chandra X-ray Observatory Center, which is operate by the Smithsonian Astrophysical Observatory for and on the behalf of the National Aeronautics Space Administration under contract NAS8-03060. JH would like to thank George Younes for useful discussions regarding this paper. We would like
to thank the anonymous referee for careful reading of the paper and useful comments.

\end{document}